\pdfoutput=1
\documentclass[11pt]{article}
\usepackage{amsmath}
\usepackage{subcaption}
\usepackage{booktabs}
\usepackage{graphicx}
\usepackage{tabularx}
\usepackage{xcolor}
\usepackage{indentfirst}
\usepackage[a4paper,margin=2.0cm]{geometry}
\usepackage[colorlinks=true,
            linkcolor=blue!55!black,
            citecolor=blue!55!black,
            urlcolor=blue!55!black,
            pdfborder={0 0 0}]{hyperref}
\usepackage{titlesec}
\usepackage{microtype}
\renewcommand{\thesection}{\Roman{section}}
\renewcommand{\thesubsection}{\alph{subsection}}
\titleformat{\section}[block]{\normalfont\normalsize\bfseries\centering}{\thesection.}{0.45em}{}
\titleformat{\subsection}[block]{\normalfont\small\bfseries\raggedright}{\thesubsection.}{0.4em}{}

\renewenvironment{abstract}{%
  \par\begin{center}\bfseries Abstract\end{center}\small\noindent
}{\par\normalsize}

\begin{document}

\title{Low-Latency Beam-Background Suppression for the BESIII Trigger System Using Machine Learning Trained on Experimental Data}
\author{Hanlin Zhang\textsuperscript{1}, Minggang Zhao\textsuperscript{1}, and Junhao Yin\textsuperscript{1,\textdagger}\\
Boping Chen\textsuperscript{2,3}, Jingzhou Zhao\textsuperscript{2,3}, Wenxuan Gong\textsuperscript{2,3}, and Ke Li\textsuperscript{2,3,\textdagger}\\[3pt]
\normalsize \textsuperscript{1}Nankai University, Tianjin 300071, China\\
\normalsize \textsuperscript{2}Institute of High Energy Physics, Chinese Academy of Sciences, Beijing 100049, China\\
\normalsize \textsuperscript{3}University of Chinese Academy of Sciences, Beijing 100049, China\\
}
\date{}

\twocolumn[
\maketitle
\begin{abstract}
Beam-induced backgrounds contribute to the trigger and data-acquisition rate at high-luminosity electron--positron colliders. Their dependence on beam and machine conditions motivates data-driven studies alongside simulation. 
A data-driven machine-learning method for trigger-level beam-background suppression is developed using experimental data recorded by BESIII.
Compact neural-network classifiers are trained and evaluated for two trigger configurations: one based only on tracking and time-of-flight information, and one based only on electromagnetic-calorimeter information.
Each classifier uses 48 binary Level-1 trigger conditions of the BESIII detector.
The models distinguish hadron events from beam-background events and are evaluated against the established trigger selection at matched hadron-event efficiency. 
Relative to the established trigger selection, the beam-background false-positive rate is reduced by 20--60\% across the studied configurations and center-of-mass energies, with stable suppression performance as tested using random-trigger data.
Hardware-oriented synthesis indicates that the compact network is suitable for low-latency FPGA implementation. The proposed method can reduce the fraction of beam-background events in the trigger stream in future BESIII trigger upgrades. It also provides a reference for trigger optimization at future high-luminosity electron--positron colliders.
\end{abstract}
\par\medskip
]

\begingroup
\renewcommand{\thefootnote}{\fnsymbol{footnote}}

\begingroup
\renewcommand{\thefootnote}{\fnsymbol{footnote}}

\makeatletter
\long\def\@makefntext#1{%
  \noindent
  \begin{tabular}[t]{@{}p{0.7em}@{\hspace{0.5em}}p{\dimexpr\linewidth-2em\relax}@{}}
    \@makefnmark & #1
  \end{tabular}%
}
\makeatother

\footnotetext[2]{%
Corresponding authors:\par
Junhao Yin (\href{mailto:yinjh@nankai.edu.cn}{yinjh@nankai.edu.cn})\par
Ke Li (\href{mailto:like@ihep.ac.cn}{like@ihep.ac.cn})%
}
\endgroup
\endgroup

\section{INTRODUCTION}
\label{sec:introduction}
High-luminosity \(e^+e^-\) colliders offer a cleaner collision environment and a much higher signal-to-background ratio than hadron colliders. Their trigger systems can therefore retain a broad range of physics processes, including low-multiplicity and otherwise subtle final states, rather than selecting only a small set of rare signatures. Interactions involving beam particles, residual gas, beam losses, and accelerator components can generate detector activity unrelated to \(e^+e^-\) collisions. At sufficiently high luminosity, these backgrounds can constitute a substantial fraction of the low-latency event stream and motivate continued optimization of front-end buffers, DAQ, reconstruction, and analysis resources.

Machine-learning methods are increasingly being explored to optimize trigger systems in collider experiments. FPGA tools now make low-latency inference practical~\cite{atlas_muon_nn,fastml_hls4ml,Duarte:2018ite,Schulte:2025mai}. The appropriate optimization target depends on the collision environment. At hadron colliders, the large inelastic collision rate and complex event environment require trigger systems to select a limited set of signal-like signatures from a much larger event stream. Dedicated tagging and event-cleaning tools then handle beam-induced and other non-collision backgrounds. For example, ATLAS used background-enriched data from unpaired proton bunches to develop and validate beam-induced-background tags. The tags used observables from the inner detector, muon spectrometer, and calorimeters~\cite{atlas_bib_tagging}. In contrast, an \(e^+e^-\) trigger can retain an inclusive physics programme. Rejecting the dominant machine-related background then becomes an important additional trigger decision. The classifier studied here follows this background-veto logic rather than a signal-tagging logic for a particular physics process.

This background-veto design will become increasingly important for future high-luminosity lepton colliders, including the STCF, CEPC, and FCC-ee~\cite{stcf_l1_trigger,cepc_cdr,fcc_ee_physics}. Beam-induced backgrounds depend strongly on time-dependent beam conditions, machine parameters, vacuum conditions, and local accelerator losses. Their detailed detector response can therefore benefit from experimental data in addition to Monte Carlo simulation. This motivates approaches that learn directly from experimental data.

The Beijing Electron-Positron Collider II (BEPCII)~\cite{bePCII}
and the Beijing Spectrometer III (BESIII)~\cite{besIII_detector}
provide an operating platform for studies using real data.
BEPCII is a double-ring \(e^+e^-\) collider in the \(\tau\)-charm energy region. Its peak luminosity is \(1\times10^{33}~\mathrm{cm^{-2}\,s^{-1}}\) at \(1.89~\mathrm{GeV}\). The BEPCII high-energy upgrade equipment installation was completed in 2024, and the peak luminosity after the upgrade is expected to increase to about three times at \(2.35~\mathrm{GeV}\)~\cite{bepcII_upgrade}.
The BESIII detector records \(e^+e^-\) collision data delivered by
the BEPCII storage ring and has produced a broad range of important
results in hadron spectroscopy, charm physics, baryon decays, and
precision measurements. Representative examples include the discovery of the charged
charmonium-like states $Z_c(3900)$ and $Z_c(4020)$
\cite{BESIII_Zc3900,BESIII_Zc4020}, the observation of the $X(3872)$
in $e^+e^-$ annihilation \cite{BESIII_X3872}, precision measurements
of absolute $\Lambda_c^+$ branching fractions
\cite{BESIII_LambdaC} and strong-phase parameters in quantum-correlated
$D^0\bar{D}^0$ decays \cite{BESIII_DStrongPhase}, measurements of the
$R$ value in $e^+e^-\to\mathrm{hadrons}$ annihilation
\cite{BESIII_RValue}, the observation of the excited
$\Omega(2012)^-$ baryon \cite{BESIII_Omega2012}, and measurements of
hyperon decay parameters and tests of CP symmetry
\cite{BESIII_LambdaCP}.
At the \(J/\psi\) peak, \(\sqrt{s}=3.097~\mathrm{GeV}\),
beam-background events that pass loose Level-1 (L1) requirements
occur at approximately the same rate as hadronic
events~\cite{besIII_trigger_eff_2018}. Studies at higher energy
points indicate higher beam-background levels. BEPCII plans to adopt a crab-waist beam scheme after
2030. The scheme is expected to increase the peak luminosity at
\(1.89~\mathrm{GeV}\) by a factor of ten and will make greater
demands on the trigger system~\cite{bepcii_crab_waist_feasibility}.

The BESIII trigger system~\cite{besIII_trigger_system} has successfully supported the experiment's broad
physics programme through an established configuration based on validated
detector-level information. Building on this foundation, the present study
investigates whether compact neural networks can provide an additional and
complementary approach to trigger optimization. The strategy identifies and
rejects beam-induced background while retaining physics events with high
efficiency, without targeting a particular signal signature. Neural networks
are trained and evaluated using experimental BESIII data at multiple
center-of-mass energies. Their performance is compared with the established
trigger selection, and their feasibility for FPGA deployment is assessed.
This study explores the optimization space offered by machine learning and
provides a reference for future trigger upgrades at BESIII and other
high-luminosity lepton colliders.

\section{BESIII TRIGGER SYSTEM AND DATA SAMPLES}
\label{sec:trigger_data}
The L1 trigger of BESIII integrates information from four subdetectors: the Main Drift Chamber (MDC), the Time-of-Flight (TOF) system, the Electromagnetic Calorimeter (EMC), and the Muon Counter (MUC). Their signals are converted into trigger-level quantities and evaluated by the Global Trigger Logic(GTL). The resulting decision is sent to the fast-control system. The system is synchronized with the accelerator radio frequency, and its L1 logic is implemented primarily in FPGAs. The GTL combines the subsystem information into 48 trigger conditions. These are subsequently mapped to 16 trigger channels~\cite{besIII_trigger_eff_2018,besIII_trigger_system}.

To study trigger efficiencies, two data samples were collected at the \(\sqrt{s}=3.097~\mathrm{GeV}\) resonance in 2009. The samples use single-channel configurations: one with only trigger channel 03 (CH03) enabled and one with only trigger channel 12 (CH12) enabled. The CH03 sample, denoted as ``MDC\&TOF only'', requires at least two Barrel TOF hits and at least two long tracks reconstructed in the MDC. The CH12 sample, denoted as ``EMC only'', requires at least two EMC clusters and a total energy above the middle threshold. These configurations minimize correlations between subdetector responses and allow the efficiency of each detector-based trigger channel to be evaluated independently. The samples constitute the primary dataset for this study.

To extend the evaluation to higher energies, additional data were collected in 2026 at \(\sqrt{s}=4.680~\mathrm{GeV}\) using the same single-channel configurations. These samples are referred to as ``MDC\&TOF only Data2'' and ``EMC only Data2'', respectively. Random-trigger data collected with trigger channel 10 are also used to validate trigger performance under unbiased conditions. Random-trigger samples at four center-of-mass energies assess the robustness of beam-background identification across collision energies. Table~\ref{tab:data_samples} summarizes all samples and their event counts.

\begin{table}[htbp]
\centering
\caption{Real experimental data samples used for model development and validation.}
\label{tab:data_samples}
\resizebox{\columnwidth}{!}{%
\begin{tabular}{ccc}
\toprule
 \textbf{Data sample} & \textbf{$\sqrt{s}$ (GeV)} & \textbf{Number of events}\\
\midrule
 MDC\&TOF only Data1 & 3.097 & $5.66\times10^{6}$\\
 EMC only Data1 & 3.097 & $3.06\times10^{6}$\\
\midrule
 MDC\&TOF only Data2 & 4.680 & $3.14\times10^{6}$\\
 EMC only Data2 & 4.680 & $9.28\times10^{6}$\\
\midrule
 Random trigger Data1 & 3.097 & $5.00\times10^{6}$\\
 Random trigger Data2 & 3.682 & $5.00\times10^{6}$\\
 Random trigger Data3 & 3.686 & $5.00\times10^{6}$\\
 Random trigger Data4 & 4.740 & $5.00\times10^{6}$\\
\bottomrule
\end{tabular}
}
\end{table}

\section{TRADITIONAL METHOD}
\label{sec:traditional_selection}
In the traditional method, events are classified into four categories: hadron, Bhabha, dimuon, and beam background. Explicit selection criteria define the first three categories, and events that fail all of these criteria are assigned to the beam-background class. Channel efficiencies are evaluated using Eq.~(\ref{eq:channel_efficiency}). The procedure reproduces the channel efficiencies reported in the published BESIII study with good overall agreement, and the reproduced selection therefore provides a validated, data-driven reference for the subsequent comparisons. The selection criteria are summarized in Table~\ref{tab:traditional_selection_criteria}.
\begin{table}[htbp]
\centering
\caption{Event-selection criteria used in the traditional method.}
\label{tab:traditional_selection_criteria}
\small
\begin{tabularx}{\columnwidth}{c X}
\toprule
\textbf{Event class} & \textbf{Selection criteria} \\
\midrule
Hadron & $N_{\text{good}}\ge2$; \\
       & If $N_{\text{good}}=2$, $\theta_{12}<170^\circ$. \\
Bhabha & $N_{\text{good}}=2$, $\theta_{12}>175^\circ$; \\
       & $N_{\text{clus}}=2$, $\theta_{\text{clus}}>166^\circ$; \\
       & $|E_{\text{EMC,tot}}-\sqrt{s}|/\sqrt{s} < 0.1$. \\
Dimuon & $N_{\text{good}}=2$, $\theta_{12}\ge178^\circ$; \\
       & $p_i<2$~GeV/$c$, $E_i^{\text{EMC}}<0.7$~GeV ($i=1,2$); \\
       & $E/c\in(2.8,3.3)$; \\
       & $P_x,P_y\in(-0.1,0.1)$; \\
       & $P_z\in(-0.2,0.2)$~GeV/$c$. \\
Beam background & Fails all the criteria above. \\
\bottomrule
\end{tabularx}
\medskip
\par\footnotesize ($N_{\text{good}}$ is defined by $|\cos\theta|\le0.93$, $|d_z|<10$~cm, and $|d_{xy}|<1$~cm.)
\end{table}

\begin{equation}
\epsilon_i = \frac{N_{\mathrm{ch./con.},i}}{N_{\mathrm{total}}},
\label{eq:channel_efficiency}
\end{equation}

Using this method, the trigger proportions for hadron and beam-background events are calculated for the MDC\&TOF-only and EMC-only samples. Two metrics quantify the trigger performance: the true-positive rate (TPR), defined as the fraction of hadron events accepted by the trigger, and the false-positive rate (FPR), defined as the fraction of beam-background events that pass the trigger selection. Table~\ref{tab:tpr_fpr_combined} presents the results for the individual channels.

\begin{table}[htbp]
\centering
\caption{Overall trigger proportions for hadron and beam-background events in the two data samples.}
\label{tab:tpr_fpr_combined}
\begin{tabular}{lcc}
\toprule
\textbf{Sample} & \textbf{TPR (\%)} & \textbf{FPR (\%)} \\
\midrule
MDC\&TOF only & 99.78 & 14.96 \\
EMC only & 99.93 & 70.64 \\
\bottomrule
\end{tabular}
\end{table}

\section{MACHINE-LEARNING METHOD}
\label{sec:ml_method}

For the $\sqrt{s}=3.097~\mathrm{GeV}$ dataset, separate models are trained for the two single-channel samples. Their trigger-condition distributions differ substantially. Combining them would mix distinct input distributions and complicate a direct comparison with the traditional selection. Each model uses a 6:4 stratified random split of events into training and validation samples. Binary classifiers are first trained to quantify the additional beam-background suppression achievable at matched hadron-event efficiency, followed by four-class classifiers to evaluate performance with a more realistic event composition.

The subsequent comparisons quantify the performance difference between the NN and traditional methods. The established trigger selection is a validated solution obtained by balancing the physics requirements against the available hardware bandwidth. The NN study is an exploratory effort that evaluates the potential role of neural networks as a complementary option for trigger optimization. Such studies are also relevant to the design of future higher-luminosity accelerators, where increased event rates will motivate continued trigger-system development.

\subsection{Model Architecture and Decision Thresholds}

 To satisfy the latency and FPGA-resource constraints of the trigger system, a compact feedforward neural network with two hidden layers is employed. The network takes the 48-dimensional vector of trigger conditions as input. It returns a normalized score $S_c$ for each event class $c$, where $c$ denotes the class index. Trigger decisions compare the relevant class score, or a specified combination of class scores, with an operating threshold $\tau$. The threshold and any score-combination parameters are selected on the validation sample and chosen to reproduce or exceed the relevant traditional-trigger efficiency while minimizing the beam-background FPR.

\subsection{Binary-Classification Models}
\label{sec:binary_baseline_and_hadron}

\begin{figure}[htbp]
    \centering
    \begin{subfigure}[t]{\columnwidth}
        \centering
        \includegraphics[width=\textwidth]{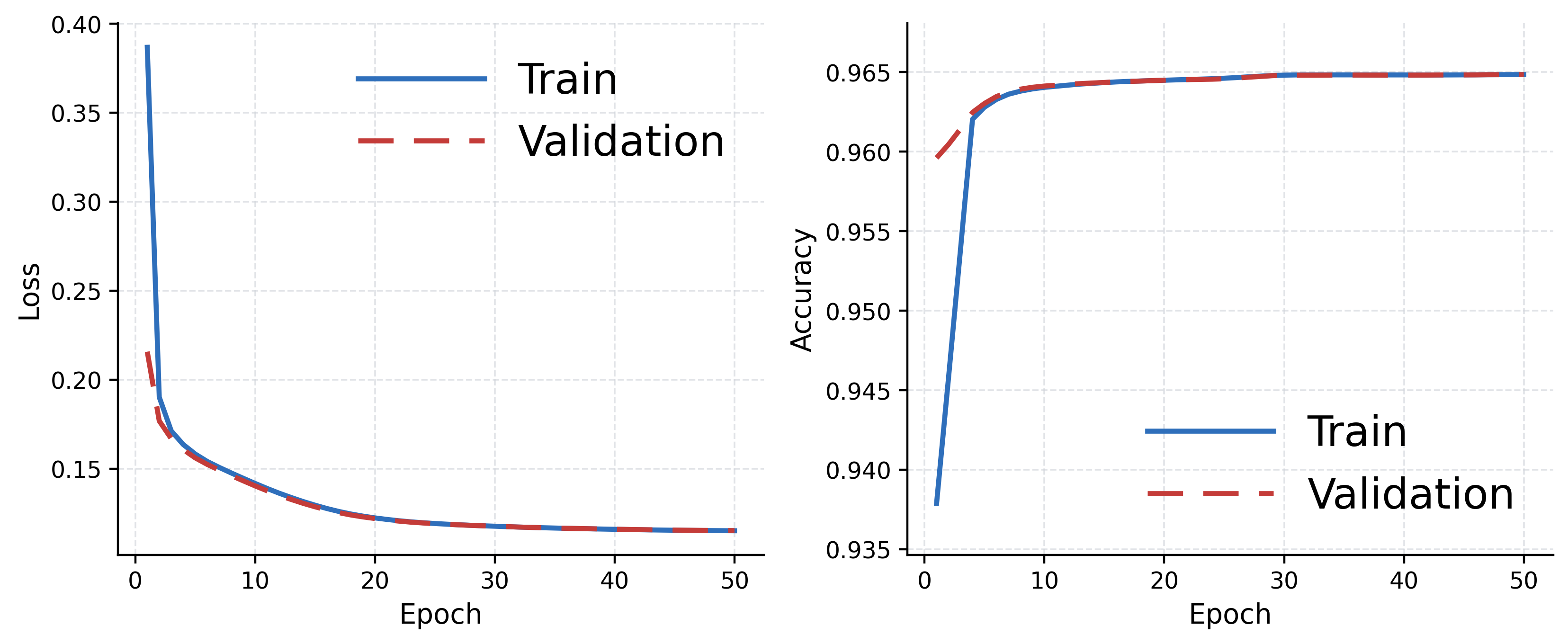}
        \caption{MDC\&TOF only}
        \label{fig:jpsi_training_mdc_tofonly}
    \end{subfigure}
    \par\medskip
    \begin{subfigure}[t]{\columnwidth}
        \centering
        \includegraphics[width=\textwidth]{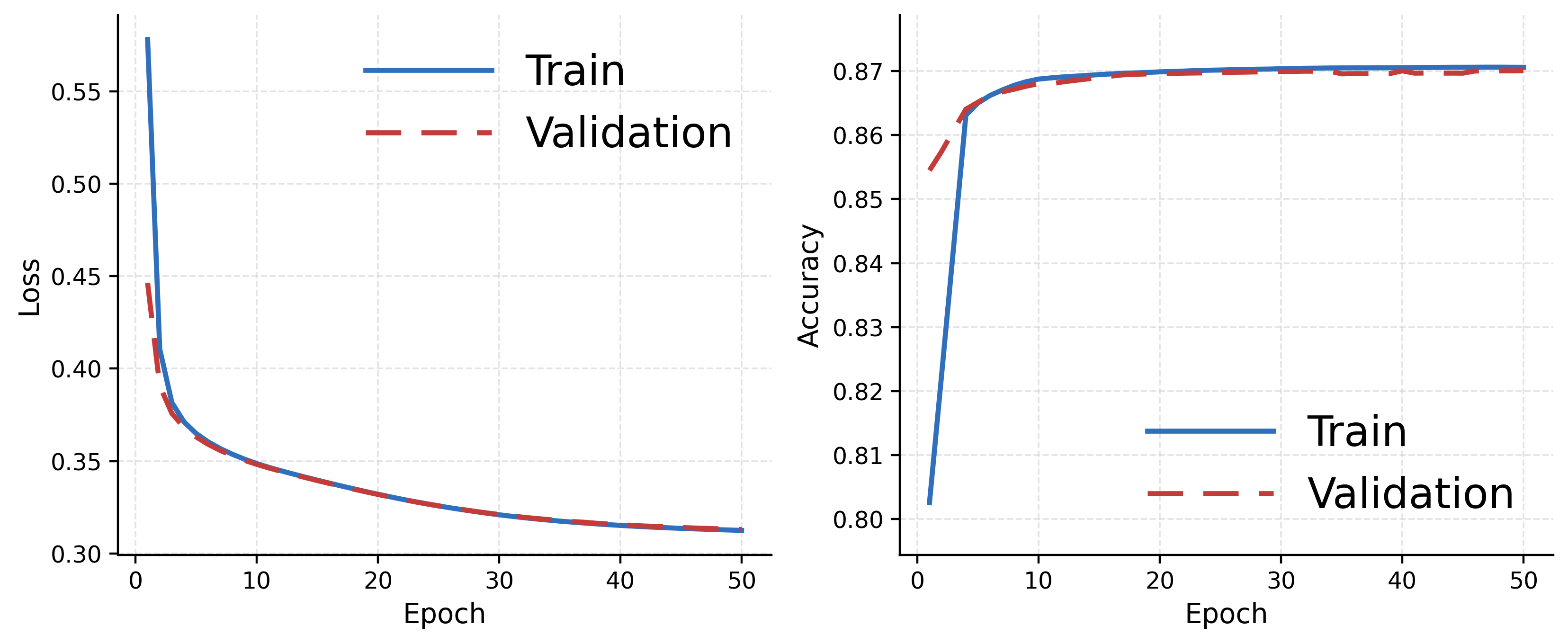}
        \caption{EMC only}
        \label{fig:jpsi_training_emconly}
    \end{subfigure}
    \caption{Training performance of the baseline binary classifier at \(\sqrt{s}=3.097~\mathrm{GeV}\).}
    \label{fig:jpsi_training_performance}
\end{figure}

A baseline binary classifier is first trained to distinguish signal events from background events. The background class comprises only beam-background events, whereas the signal class combines all events tagged as Hadron, Bhabha, or Dimuon. During training, the loss and accuracy on the training and validation samples are monitored as standard supervised-classification diagnostics~\cite{lecun_deep_learning}. The model is optimized using binary cross-entropy loss. An event is accepted as signal when $S_{\mathrm{Signal}}>\tau_{\mathrm{bin}}$. The threshold $\tau_{\mathrm{bin}}$ is selected from the receiver-operating-characteristic (ROC) curve on the validation sample according to the required operating point~\cite{fawcett_roc}. The resulting learning curves are shown in Fig.~\ref{fig:jpsi_training_performance}.

\begin{figure*}[htbp]
    \centering
    \begin{subfigure}[b]{0.42\textwidth}
        \centering
        \includegraphics[width=\textwidth]{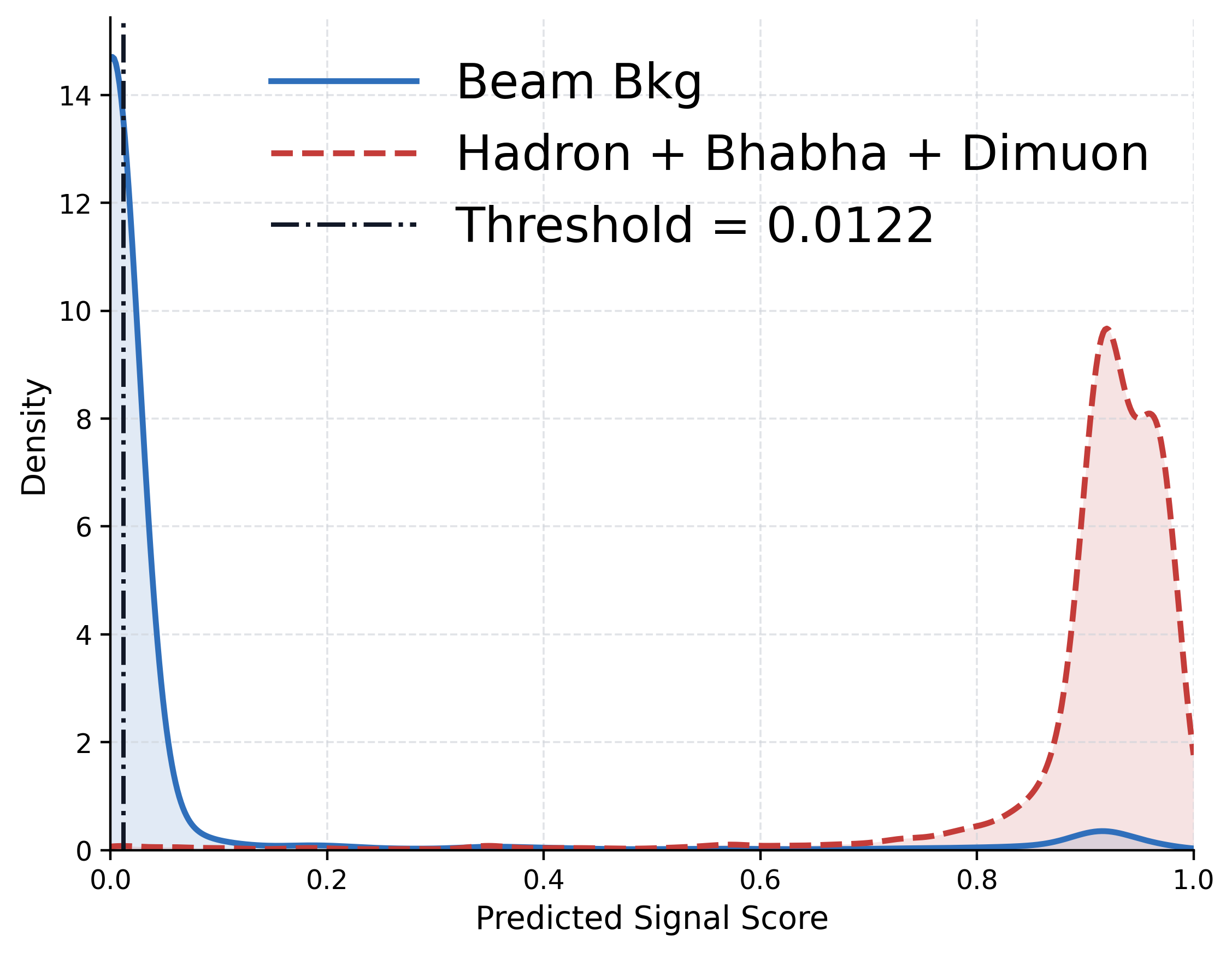}
        \caption{MDC\&TOF only}
        \label{fig:jpsi_binary_mdc_score}
    \end{subfigure}
    \hfill
    \begin{subfigure}[b]{0.42\textwidth}
        \centering
        \includegraphics[width=\textwidth]{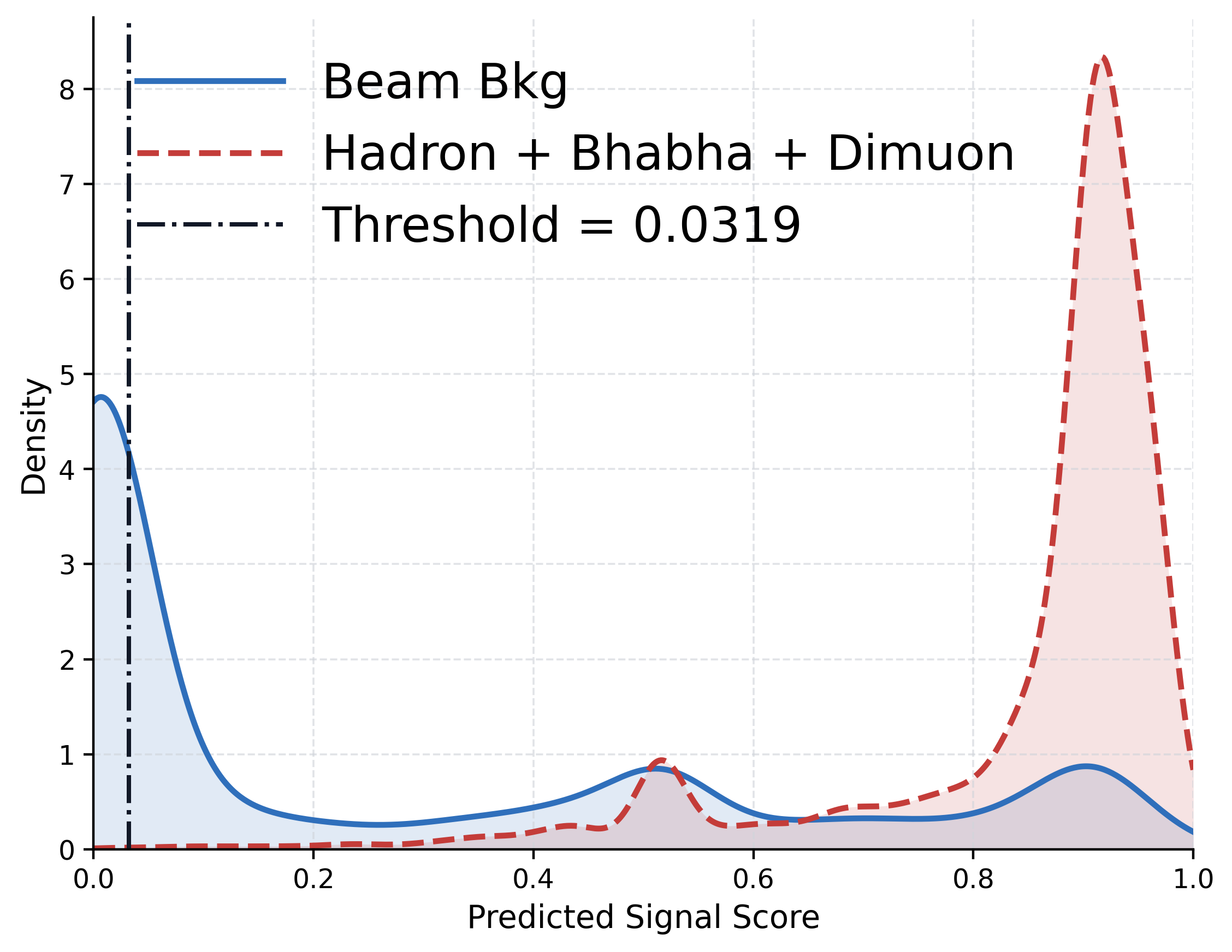}
        \caption{EMC only}
        \label{fig:jpsi_binary_emc_score}
    \end{subfigure}
    \caption{Score distributions of the baseline binary classifier. The selected threshold is indicated in each panel.}
    \label{fig:score_dists}
\end{figure*}

Training is stable for both feature configurations. The MDC\&TOF-only configuration shows greater separation than the EMC-only configuration. Figures~\ref{fig:jpsi_binary_mdc_score} and~\ref{fig:jpsi_binary_emc_score} show the baseline score distributions and the thresholds selected to satisfy the target TPR.

\begin{figure*}[htbp]
    \centering
    \begin{subfigure}[b]{0.42\textwidth}
        \centering
        \includegraphics[width=\textwidth]{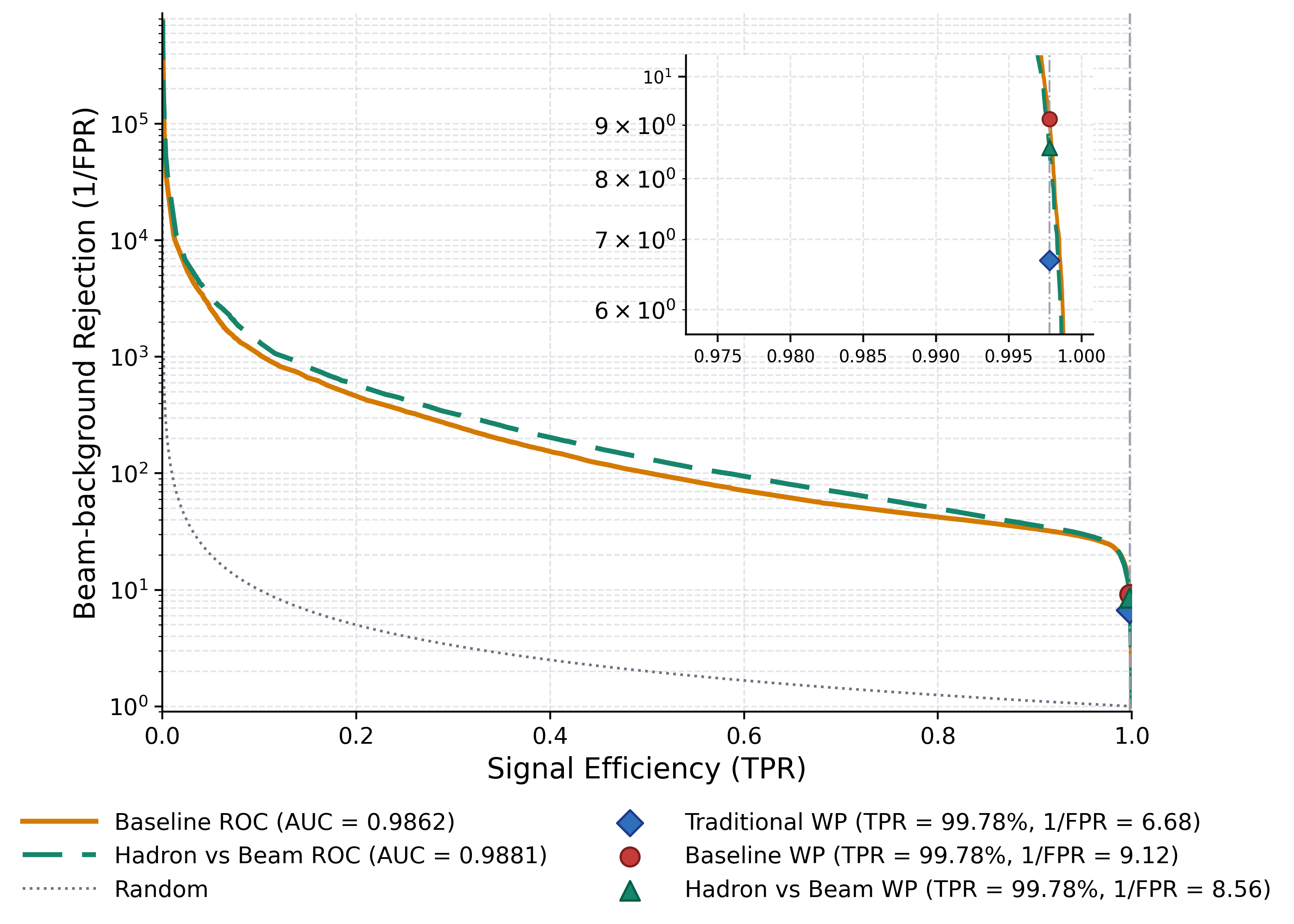}
        \caption{MDC\&TOF only}
        \label{fig:roc_binary_mdc}
    \end{subfigure}
    \hfill
    \begin{subfigure}[b]{0.42\textwidth}
        \centering
        \includegraphics[width=\textwidth]{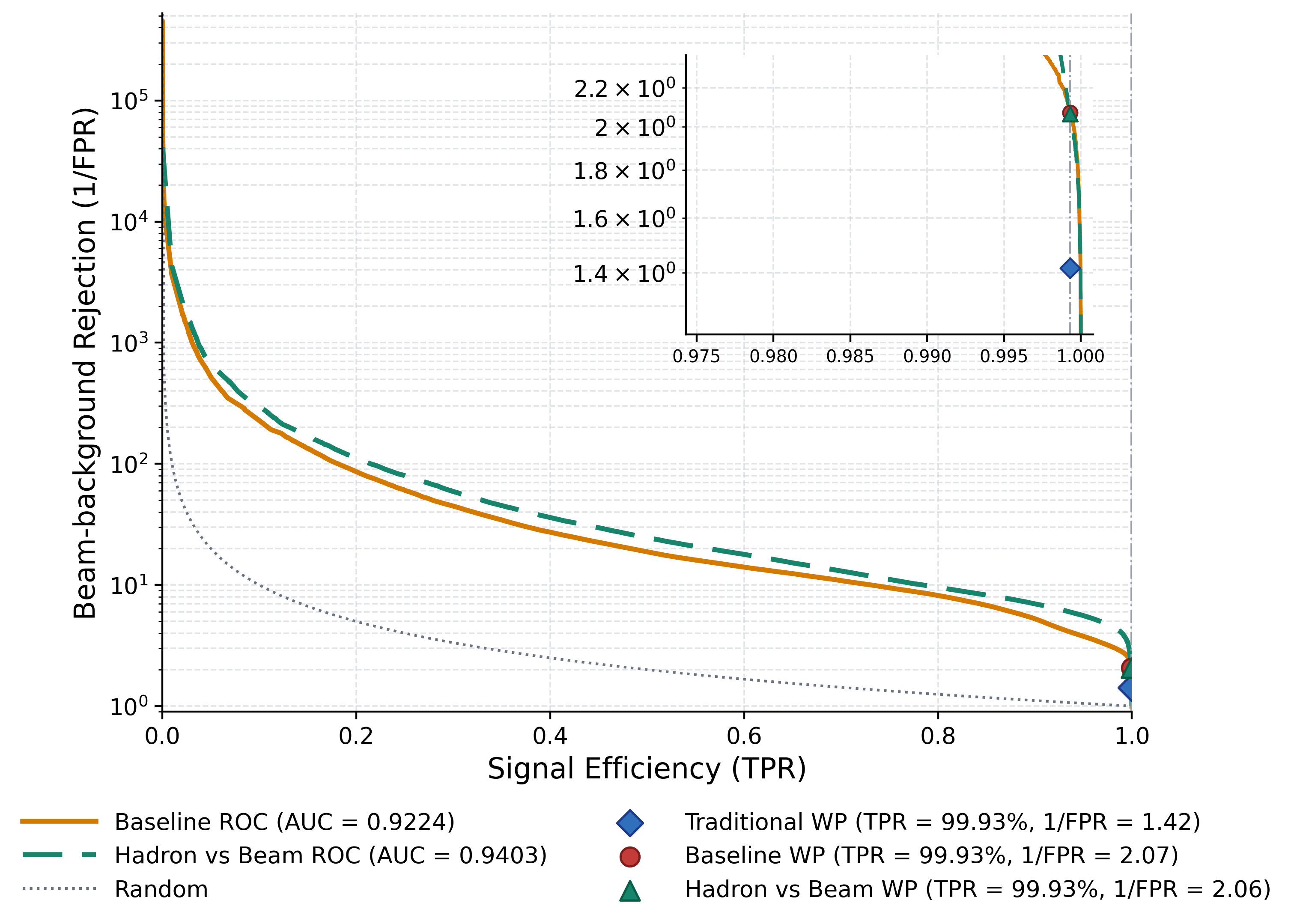}
        \caption{EMC only}
        \label{fig:roc_binary_emc}
    \end{subfigure}
    \caption{ROC curves for the baseline and Hadron-vs-Beam binary classifiers. The operating point of the traditional method is shown for comparison.}
    \label{fig:combined_roc_curves}
\end{figure*}

ROC curves are used to quantify classifier performance relative to the traditional selection. 
The relative false-positive-rate reduction ($R_{\mathrm{FPR}}$) is defined as
\begin{equation}
R_{\mathrm{FPR}} =
\frac{\mathrm{FPR}_{\mathrm{traditional}}-\mathrm{FPR}_{\mathrm{NN}}}
{\mathrm{FPR}_{\mathrm{traditional}}},
\label{eq:perf_improve}
\end{equation}
where $\mathrm{FPR}_{\mathrm{traditional}}$ and $\mathrm{FPR}_{\mathrm{NN}}$ are evaluated at matched signal-event TPR for the traditional method and the NN model, respectively. 
Figure~\ref{fig:combined_roc_curves} and Table~\ref{tab:combined_binary_perf} present the ROC curves and numerical results for the binary classifiers.
The training sample is collected under established L1 trigger requirements. It therefore contains event patterns shaped by the existing trigger logic. To complement this matched-sample comparison, the classifier is tested on random-trigger data in Section~\ref{sec:random-trigger-test}. These data provide an independent sample of beam-background events.

The baseline model provides the reference for the subsequent studies. A dedicated Hadron-vs-Beam binary classifier is then trained for a direct comparison with the four-class model introduced later. Its positive class contains Hadron events only; Bhabha and Dimuon events are excluded from the signal category. The Hadron-vs-Beam model uses the same training procedure and feature configurations as the baseline model. The narrower positive class yields a modest change in classification performance. Figure~\ref{fig:combined_roc_curves} includes its ROC curves, and Table~\ref{tab:combined_binary_perf} lists the corresponding numerical results. This dedicated model provides a consistent reference for interpreting differences between the binary and four-class classification strategies.

\begin{table}[htbp]
\centering
\caption{Performance of the binary classification models relative to the traditional method at matched signal efficiency.}
\label{tab:combined_binary_perf}
\resizebox{\columnwidth}{!}{%
\begin{tabular}{lcccc}
\toprule
\textbf{Method} & \multicolumn{2}{c}{\textbf{MDC\&TOF only}} & \multicolumn{2}{c}{\textbf{EMC only}} \\
\midrule
& \textbf{TPR(\%)} & \textbf{FPR(\%)} & \textbf{TPR(\%)} & \textbf{FPR(\%)} \\
\midrule
Traditional method & 99.78 & 14.96$\pm0.02$ & 99.93 & 70.64$\pm0.03$ \\
Baseline NN & -- & 10.97$\pm0.01$ & -- & 48.29$\pm0.04$ \\
Hadron-vs-Beam NN & -- & 11.67$\pm0.02$ & -- & 48.40$\pm0.04$ \\
\midrule
Baseline NN $R_{\mathrm{FPR}}$ & \multicolumn{2}{c}{26.67\%} & \multicolumn{2}{c}{31.64\%} \\
Hadron-vs-Beam NN $R_{\mathrm{FPR}}$ & \multicolumn{2}{c}{21.99\%} & \multicolumn{2}{c}{31.48\%} \\
\bottomrule
\end{tabular}
}
\end{table}

\subsection{Four-Class Classification Results}
\label{sec:four_class_results}
The classifier is extended to distinguish four event classes: hadron, Bhabha, dimuon, and beam background. The event categories and input representation are identical to those used by the traditional method, enabling a direct comparison. The principal objective is to separate hadron events from beam-background events. Bhabha and dimuon events are retained when they are not classified as beam background. Their mutual separation is not optimized in this study because their score distributions are more similar to those of hadron events than to those of beam-background events.

\begin{figure*}[htbp]
    \centering
    \begin{subfigure}[b]{0.42\textwidth}
        \centering
        \includegraphics[width=\textwidth]{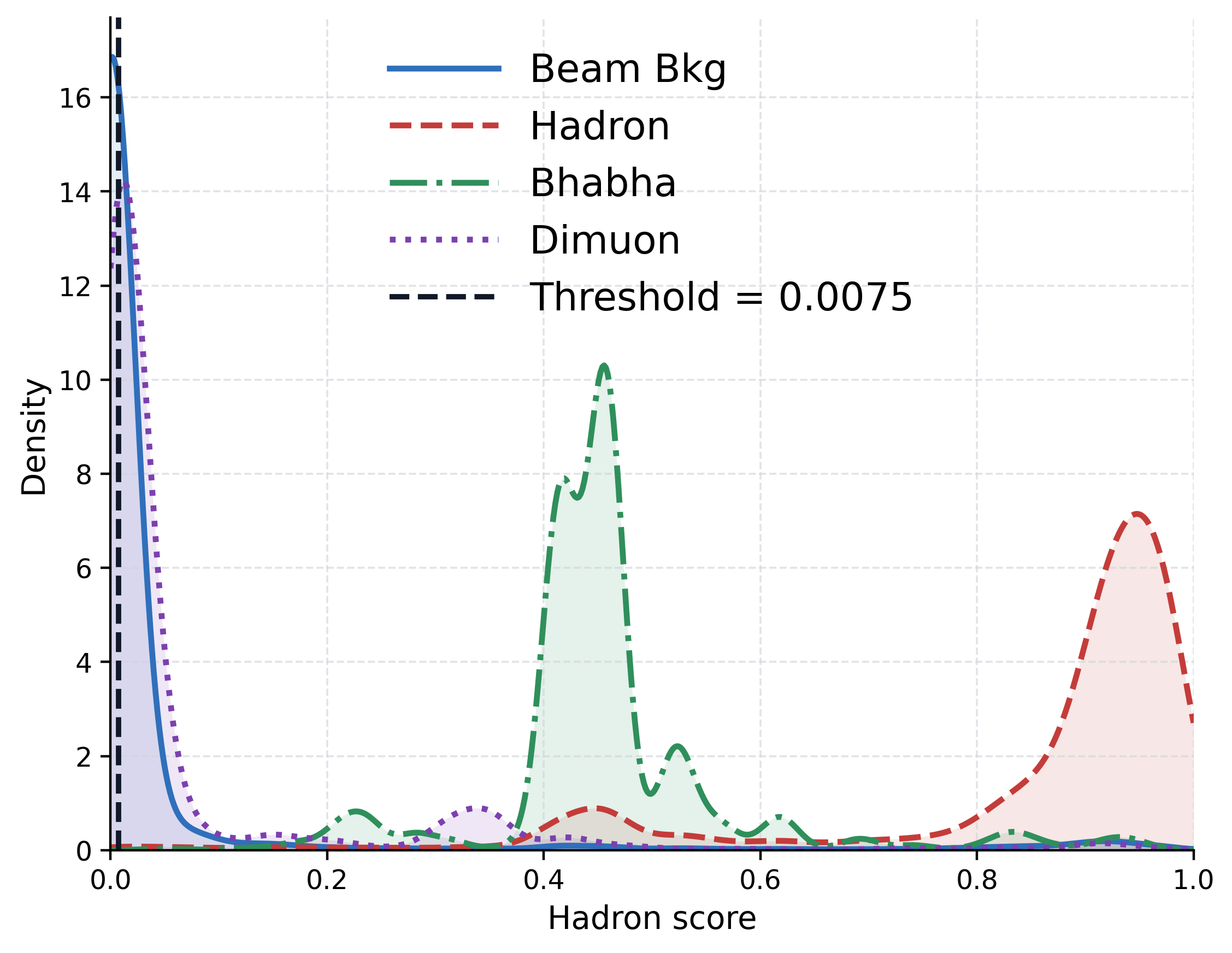}
        \caption{MDC\&TOF only: hadron score.}
        \label{fig:ch03_hadron_score}
    \end{subfigure}
    \hfill
    \begin{subfigure}[b]{0.42\textwidth}
        \centering
        \includegraphics[width=\textwidth]{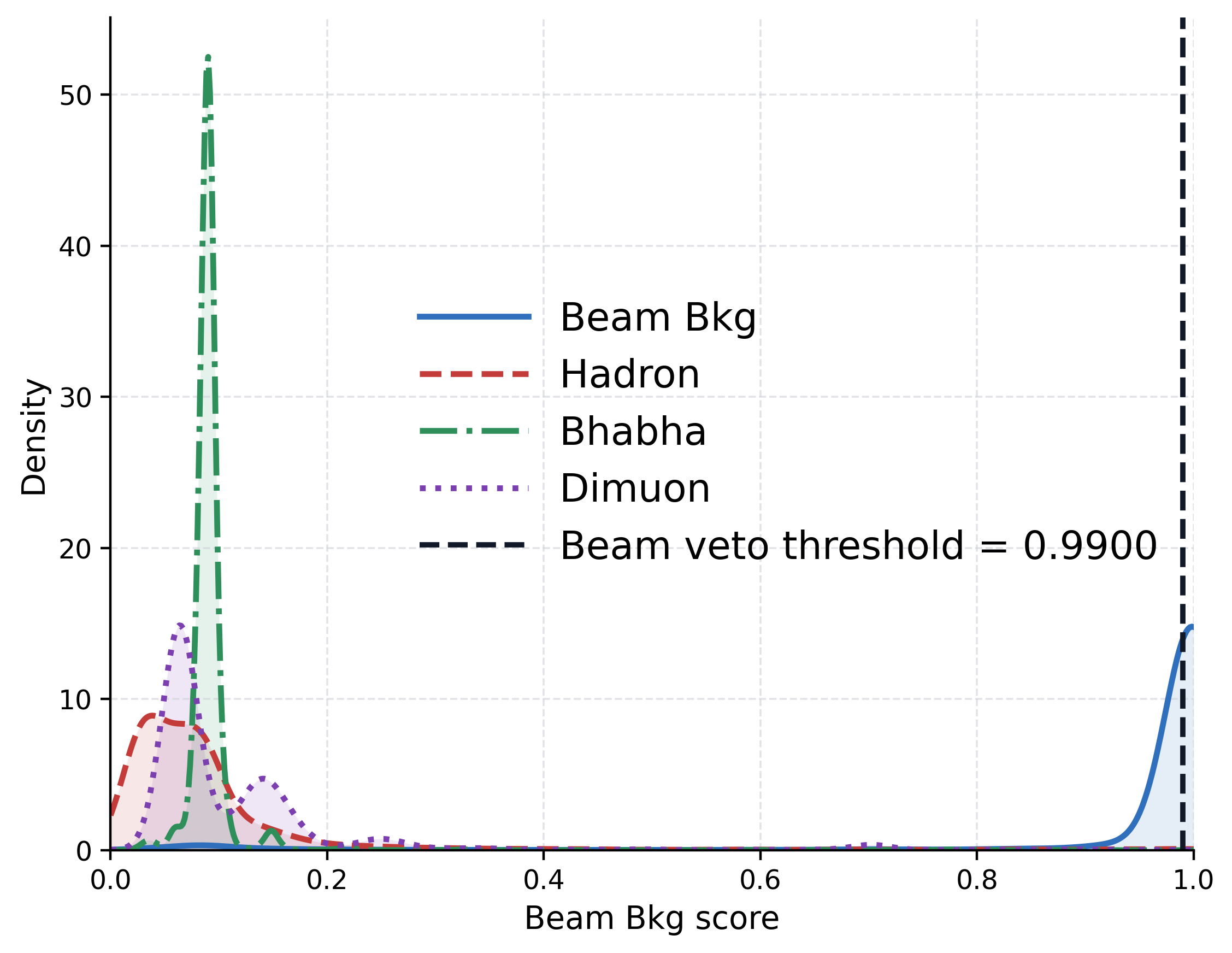}
        \caption{MDC\&TOF only: beam-background score.}
        \label{fig:ch03_beam_score}
    \end{subfigure}
    \par\medskip
    \begin{subfigure}[b]{0.42\textwidth}
        \centering
        \includegraphics[width=\textwidth]{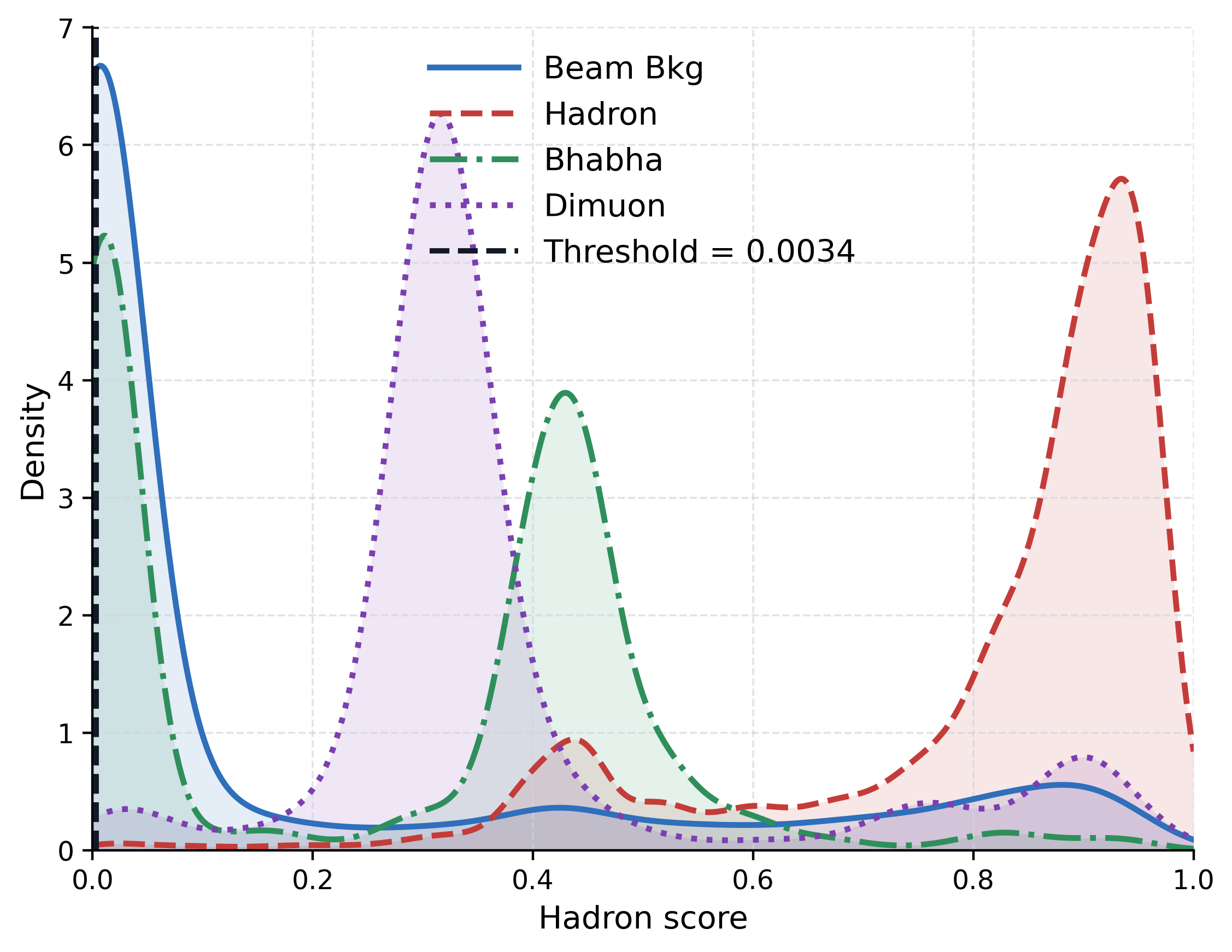}
        \caption{EMC only: hadron score.}
        \label{fig:ch12_hadron_score}
    \end{subfigure}
    \hfill
    \begin{subfigure}[b]{0.42\textwidth}
        \centering
        \includegraphics[width=\textwidth]{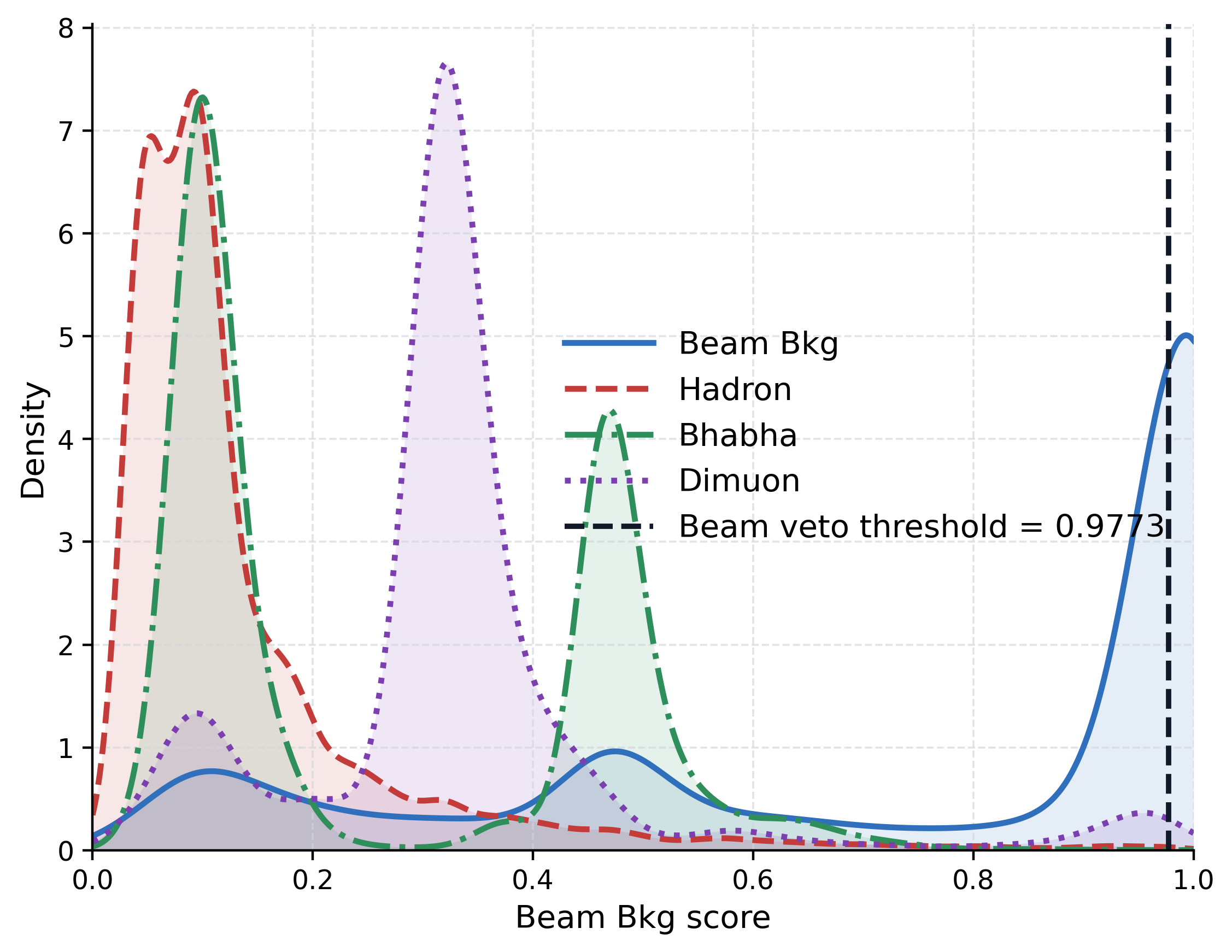}
        \caption{EMC only: beam-background score.}
        \label{fig:ch12_beam_score}
    \end{subfigure}
    \caption{Score distributions for the four-class model using the MDC\&TOF only and EMC only feature sets.}
    \label{fig:jpsi_four_score_distributions}
\end{figure*}

For the baseline four-class Hadron trigger, an event is accepted only when both requirements $S_{\mathrm{Hadron}}>\tau_{\mathrm{Hadron}}$ and $S_{\mathrm{Beam}}\leq\tau_{\mathrm{Beam}}$ are satisfied. The Hadron score must therefore exceed its threshold. The Beam score must remain below the beam-background veto threshold. The two thresholds are selected on the validation sample to achieve the required Hadron-trigger efficiency while minimizing the beam-background FPR. Figure~\ref{fig:jpsi_four_score_distributions} shows the hadron and beam-background score distributions for the two feature sets.

\begin{figure*}[htbp]
    \centering
    \begin{subfigure}[b]{0.42\textwidth}
        \centering
        \includegraphics[width=\textwidth]{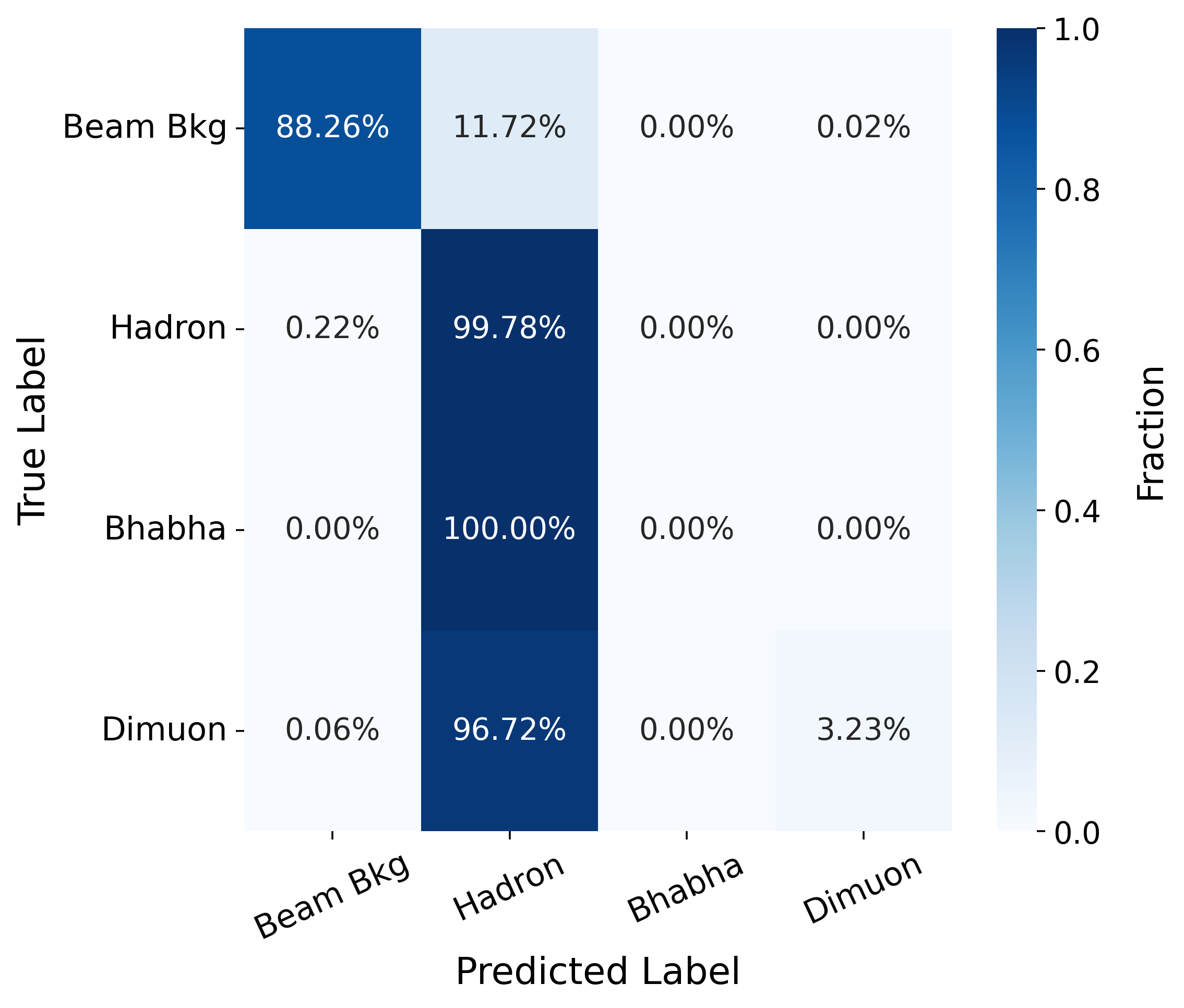}
        \caption{MDC\&TOF only}
        \label{fig:matrix_four_nn_jpsi_mdc}
    \end{subfigure}
    \hfill
    \begin{subfigure}[b]{0.42\textwidth}
        \centering
        \includegraphics[width=\textwidth]{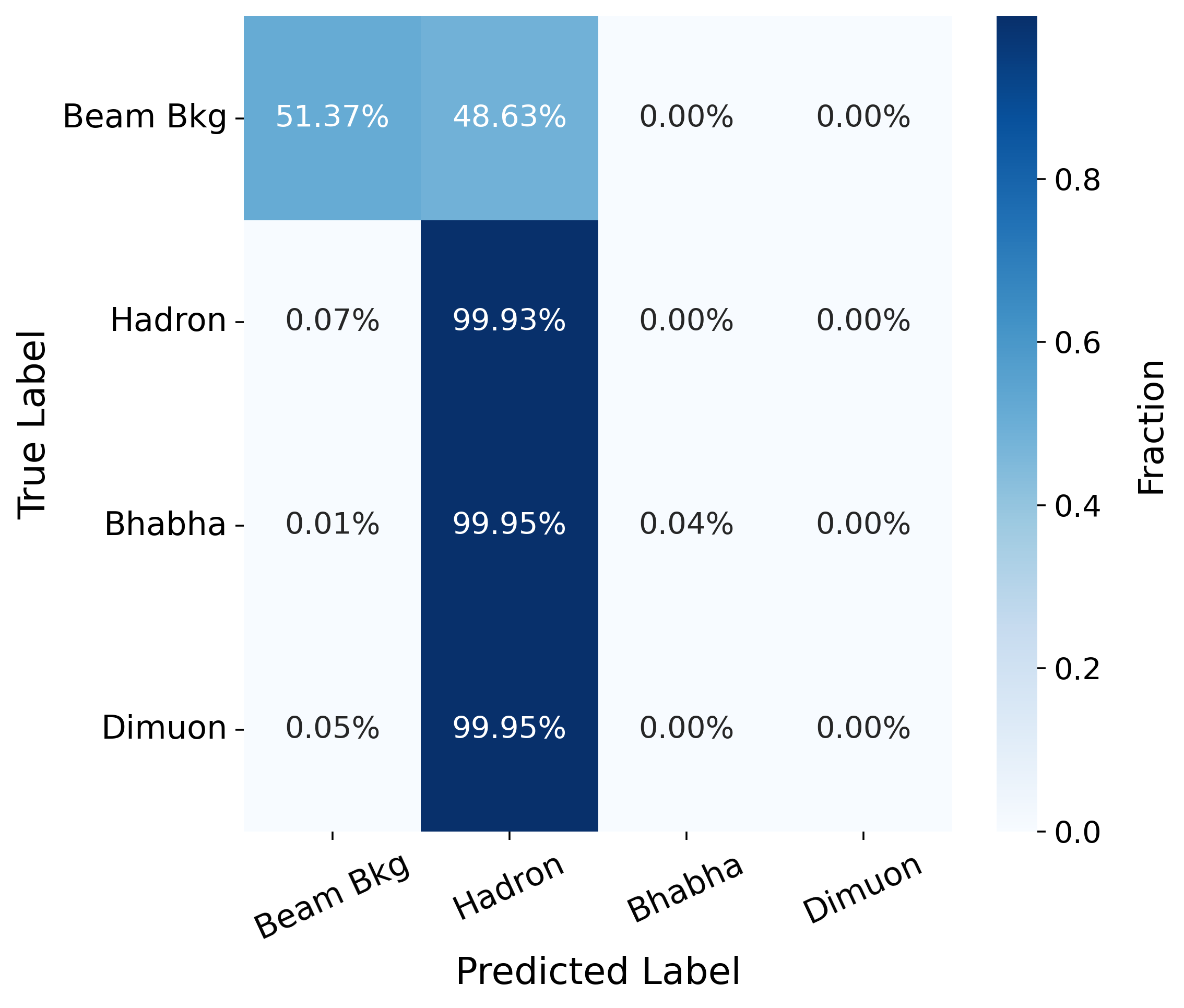}
        \caption{EMC only}
        \label{fig:matrix_four_nn_jpsi_emc}
    \end{subfigure}
    \caption{Normalized confusion matrices for the four-class models trained on the two feature sets.}
    \label{fig:jpsi_four_matrix}
\end{figure*}

Figure~\ref{fig:jpsi_four_matrix} presents the normalized confusion matrices. The FPR is calculated from beam-background events incorrectly accepted as signal. Table~\ref{tab:jpsi_four_perf_diff} summarizes the results and reports the relative FPR reduction defined in Eq.~\eqref{eq:perf_improve}.

\begin{table}[htbp]
\centering
\caption{Performance of the four-class classification model relative to the traditional method at matched hadron-event efficiency.}
\label{tab:jpsi_four_perf_diff}
\resizebox{\columnwidth}{!}{%
\begin{tabular}{ccccc}
\toprule
 \textbf{Method} & \multicolumn{2}{c}{\textbf{MDC\&TOF only}} & \multicolumn{2}{c}{\textbf{EMC only}} \\
\midrule
 & \textbf{TPR ($\%$)} & \textbf{FPR ($\%$)} & \textbf{TPR ($\%$)} & \textbf{FPR ($\%$)}\\
\midrule
  Traditional method &99.78&14.96$\pm0.02$& 99.93 & $70.64\pm0.03$\\
NN Model & - &11.74$\pm0.02$& - &$48.63\pm0.05$\\
\midrule
 $\mathbf{R}_{\mathbf{FPR}}$ & \multicolumn{2}{c}{$\mathbf{20.38\%}$} & \multicolumn{2}{c}{$\mathbf{31.16\%}$} \\
\bottomrule
\end{tabular}
}
\end{table}

\begin{figure}[htbp]
    \centering
    \includegraphics[width=\columnwidth]{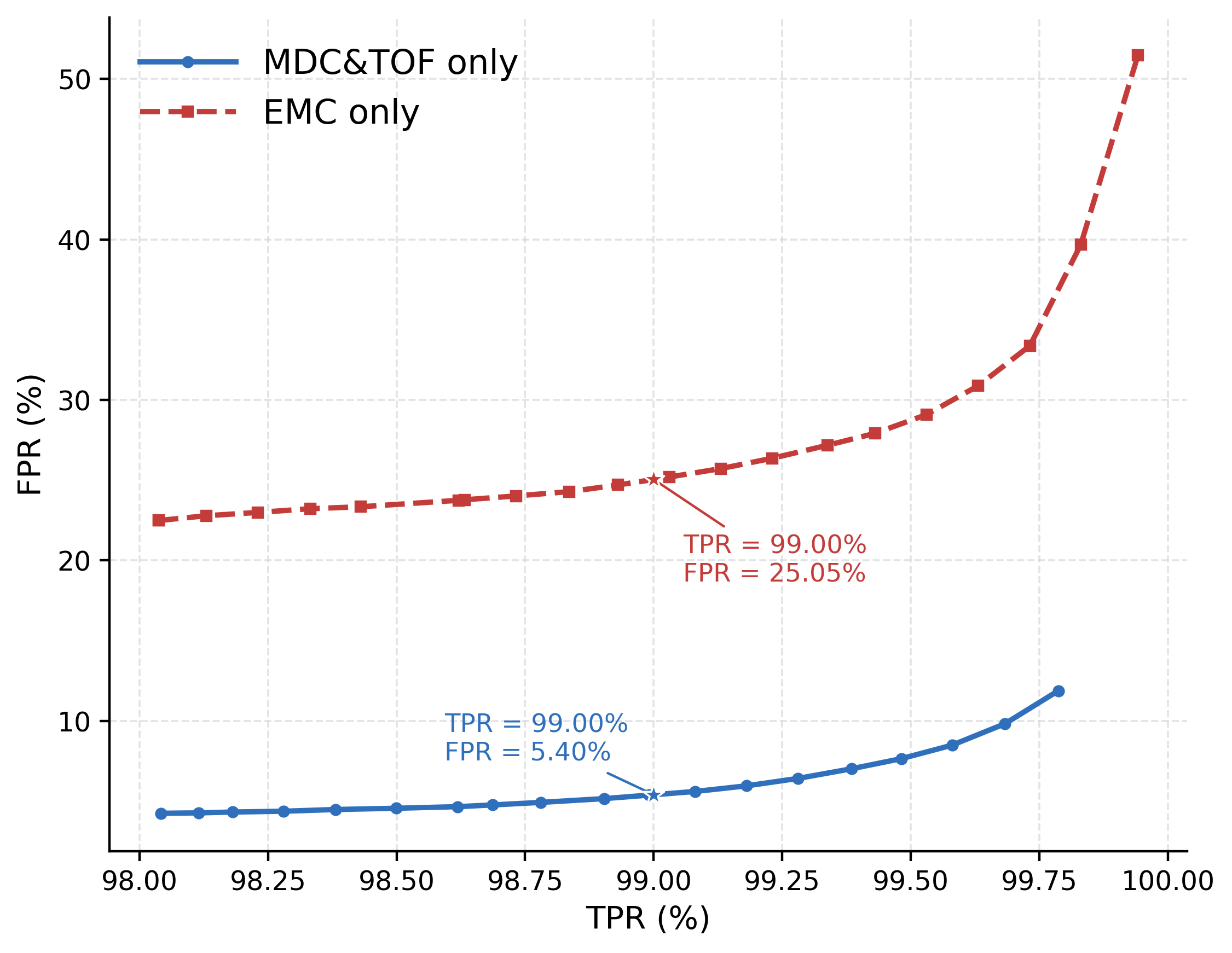}
    \caption{Beam-background FPR as a function of the hadron-event TPR for the two feature sets.}
    \label{fig:fpr_vs_tpr_all}
    
\end{figure}

Figure~\ref{fig:fpr_vs_tpr_all} shows the trade-off between TPR and FPR. A lower signal-efficiency operating point provides stronger beam-background suppression. At a fixed hadron-event TPR of 99\%, the MDC\&TOF-only and EMC-only models achieve FPRs of 5.40\% and 25.05\%, respectively.

\section{VALIDATION STUDIES}
\label{sec:validation}
\subsection{Combined-Score Validation}
\begin{figure*}[htbp]
    \centering
    \begin{subfigure}[b]{0.42\textwidth}
        \centering
        \includegraphics[width=\textwidth]{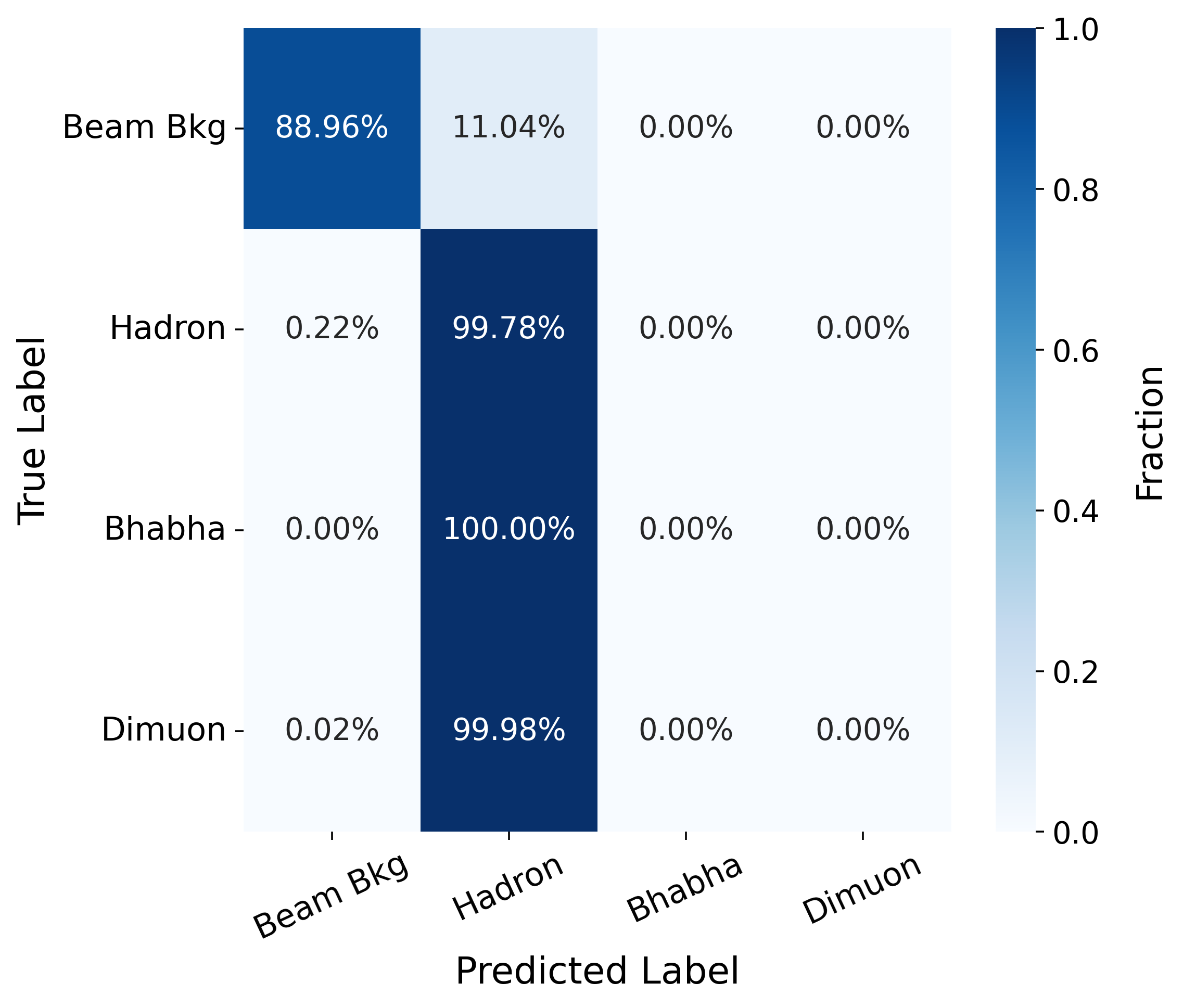}
        \caption{MDC\&TOF only}
        \label{fig:matrix_four_nn_jpsi_mdc_combine}
    \end{subfigure}
    \hfill
    \begin{subfigure}[b]{0.42\textwidth}
        \centering
        \includegraphics[width=\textwidth]{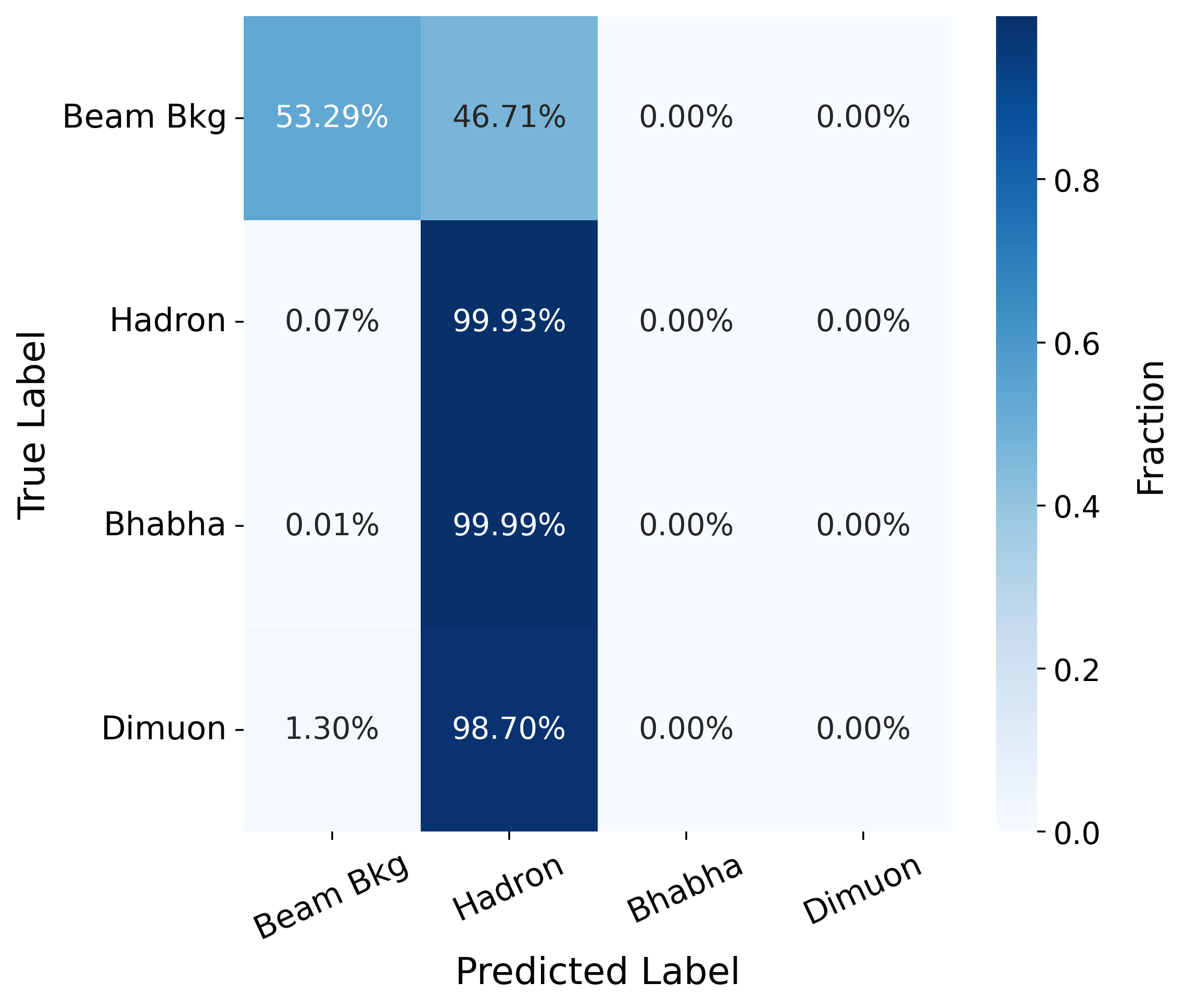}
        \caption{EMC only}
        \label{fig:matrix_four_nn_jpsi_emc_conbine}
    \end{subfigure}
    \caption{Normalized confusion matrices obtained with the combined-score single-threshold strategy.}
    \label{fig:combined_threshold_confusion_matrix}
\end{figure*}

The possibility of combining the Hadron and Beam scores into a single score, $S_{\mathrm{comb}}=S_{\mathrm{Hadron}}-\lambda S_{\mathrm{Beam}}$, is further investigated. An event is accepted when $S_{\mathrm{comb}}>\tau$. At the required hadron-trigger efficiency, this two-score combination provides comparable or slightly improved beam-background rejection relative to the double-threshold reference. It requires the evaluation of only one combined score and a single threshold comparison. In this expression, $\lambda$ is the relative Beam-score weight and $\tau$ is the operating threshold. Both parameters are determined by a scan on the threshold-selection sample and are not learned together with the neural-network parameters. The corresponding confusion matrices in Figure~\ref{fig:combined_threshold_confusion_matrix} show the performance of the combined-score strategy.

\subsection{Comparison with Boosted Decision Trees}
\begin{figure*}[htbp]
    \centering
    \begin{subfigure}[b]{0.42\textwidth}
        \centering
        \includegraphics[width=\textwidth]{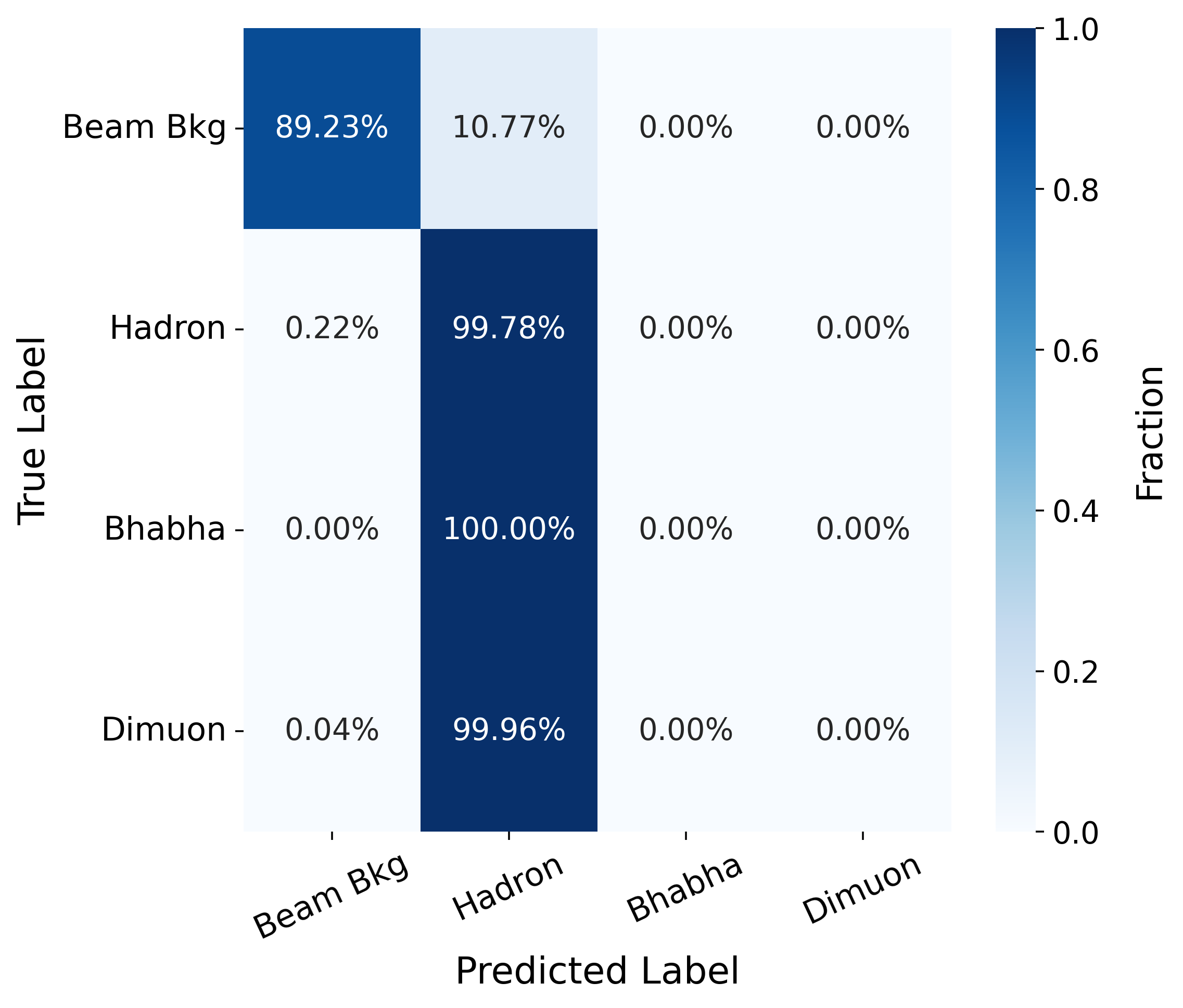}
        \caption{MDC\&TOF only}
        \label{fig:matrix_four_BDT_jpsi_mdc}
    \end{subfigure}
    \hfill
    \begin{subfigure}[b]{0.42\textwidth}
        \centering
        \includegraphics[width=\textwidth]{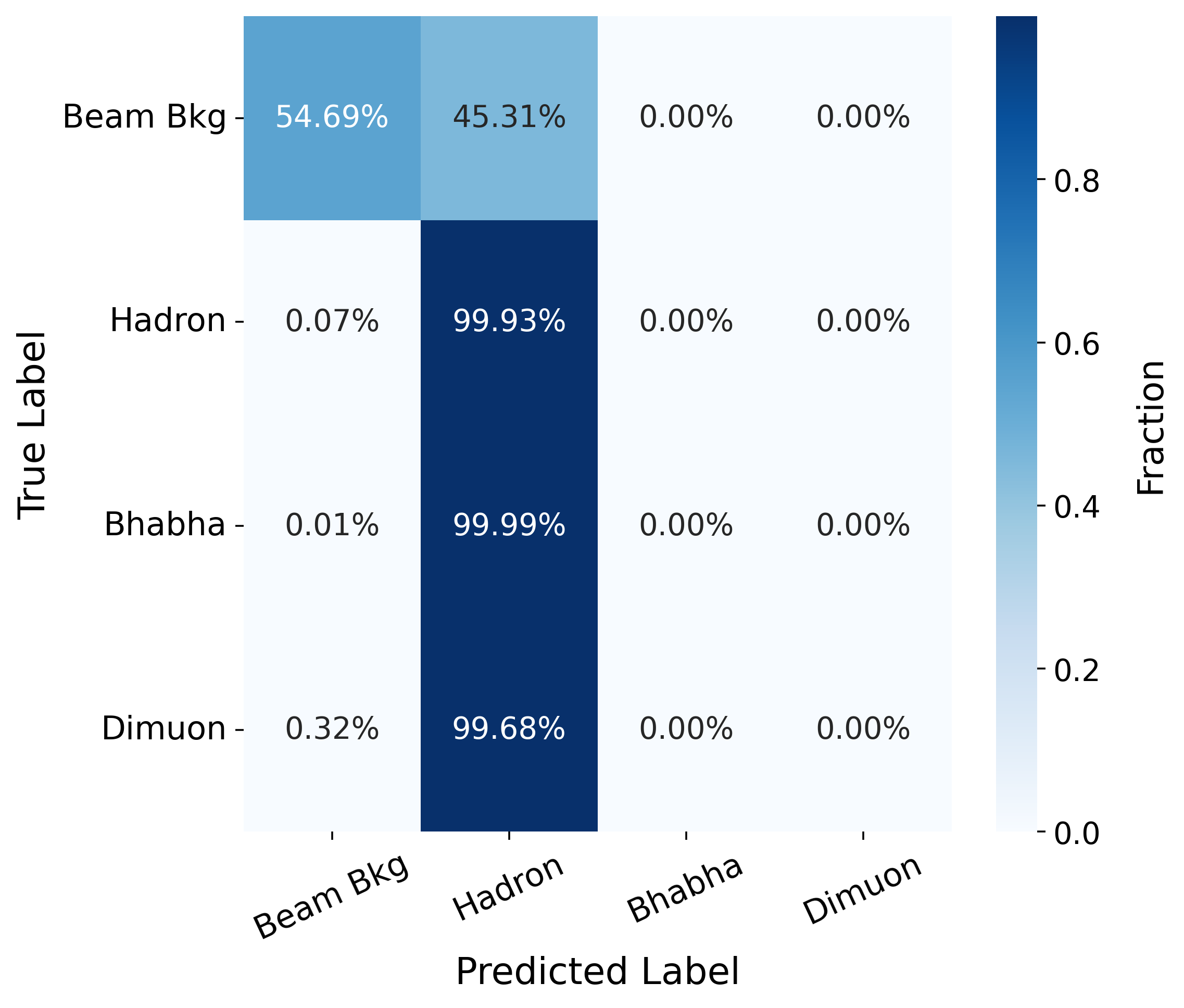}
        \caption{EMC only}
        \label{fig:matrix_four_BDT_jpsi_emc}
    \end{subfigure}
    \caption{Normalized confusion matrices of the four-class BDT model at \(\sqrt{s}=3.097~\mathrm{GeV}\) for the two feature configurations: MDC\&TOF only and EMC only.}
    \label{fig:jpsi_four_matrix_bdt}
\end{figure*}

The four-class task is also evaluated with a boosted decision tree (BDT) model. This comparison assesses whether the observed performance difference depends on the NN architecture and evaluates the NN alongside a standard BDT approach. The maximum number of BDT boosting iterations is set to 220, resulting in an ensemble with a larger capacity than the present NN architecture. The implementation requirements of both models are examined in the subsequent complexity evaluation. Figure~\ref{fig:jpsi_four_matrix_bdt} shows the BDT results for the two data samples.

\begin{figure*}[htbp]
    \centering
    \includegraphics[width=0.70\textwidth]{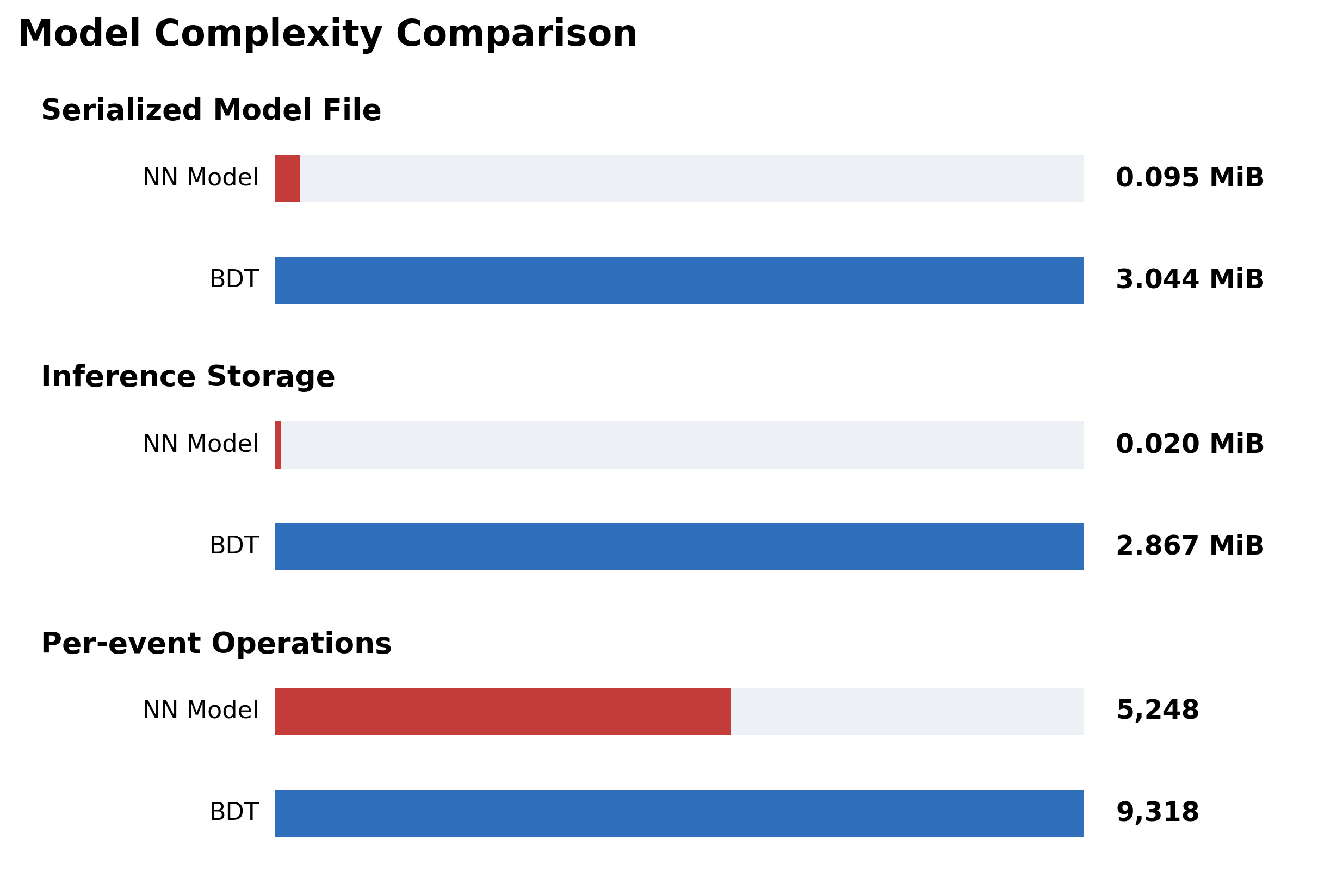}
    \caption{Comparison of the model complexity between the NN and BDT implementations.}
    \label{fig:bdt_nn_complexity}
\end{figure*}

Model complexity is subsequently evaluated for the BDT and NN implementations in Figure~\ref{fig:bdt_nn_complexity}. The comparison includes the serialized model size on disk, inference storage requirements, and the number of computational operations. FPGA inference for BDTs has been studied for particle-physics applications~\cite{bdt_fpga}. In this study, the BDT implementation requires more storage and computational operations than the NN implementation.

\subsection{Higher-Energy Performance}
At $\sqrt{s}=4.680~\mathrm{GeV}$, the architecture employed at
$\sqrt{s}=3.097~\mathrm{GeV}$ is used. Joint training with the
$\sqrt{s}=3.097~\mathrm{GeV}$ sample is also investigated. In this case, the center-of-mass energy
is an additional input feature. The resulting model is evaluated on the
$\sqrt{s}=4.680~\mathrm{GeV}$ sample.

\begin{figure*}[htbp]
\centering
\begin{subfigure}[b]{0.42\textwidth}
\centering
\includegraphics[width=\textwidth]{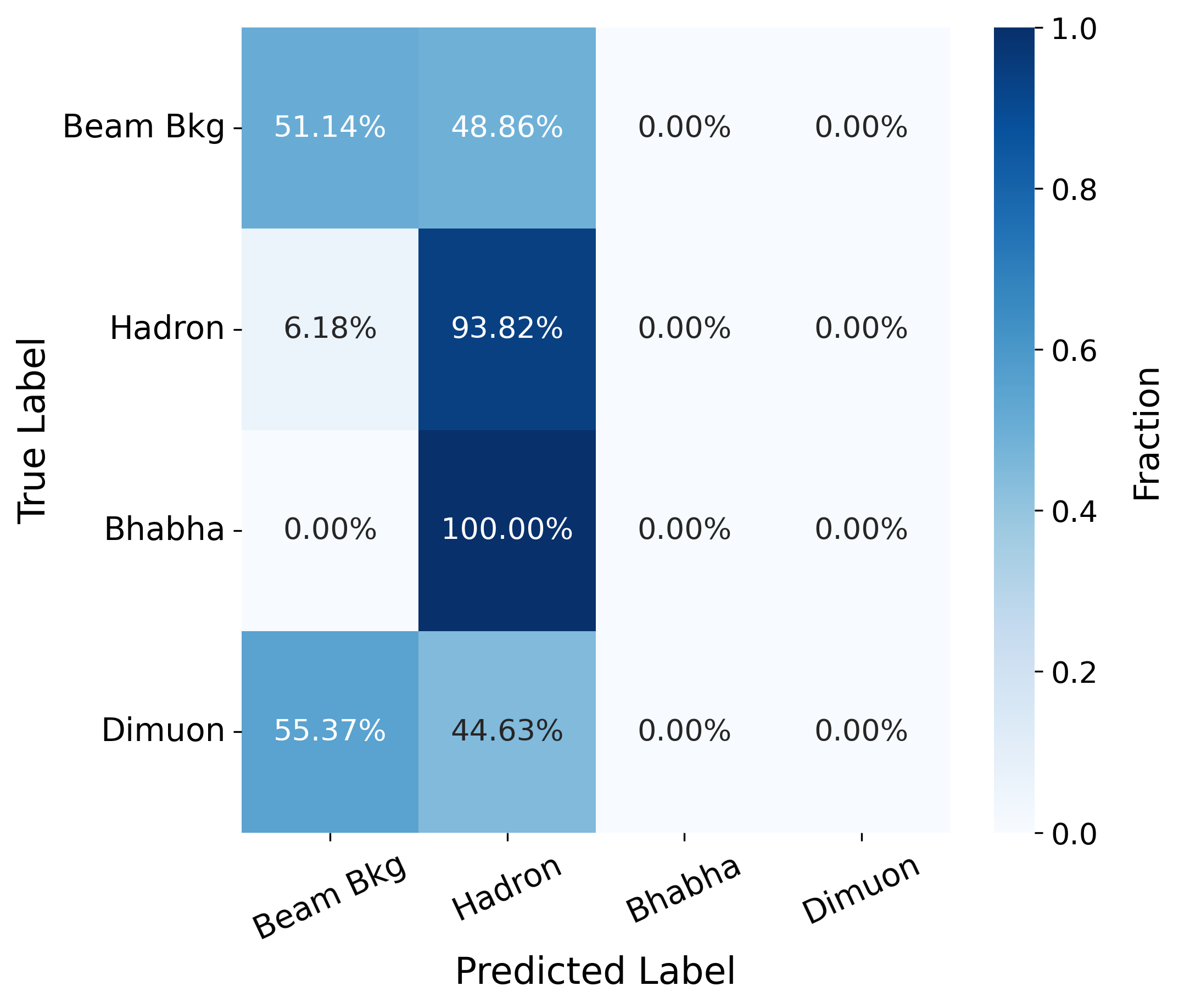}
\caption{MDC\&TOF only}
\label{fig:matrix_four_nn_4680_mdc}
\end{subfigure}
\hfill
\begin{subfigure}[b]{0.42\textwidth}
\centering
\includegraphics[width=\textwidth]{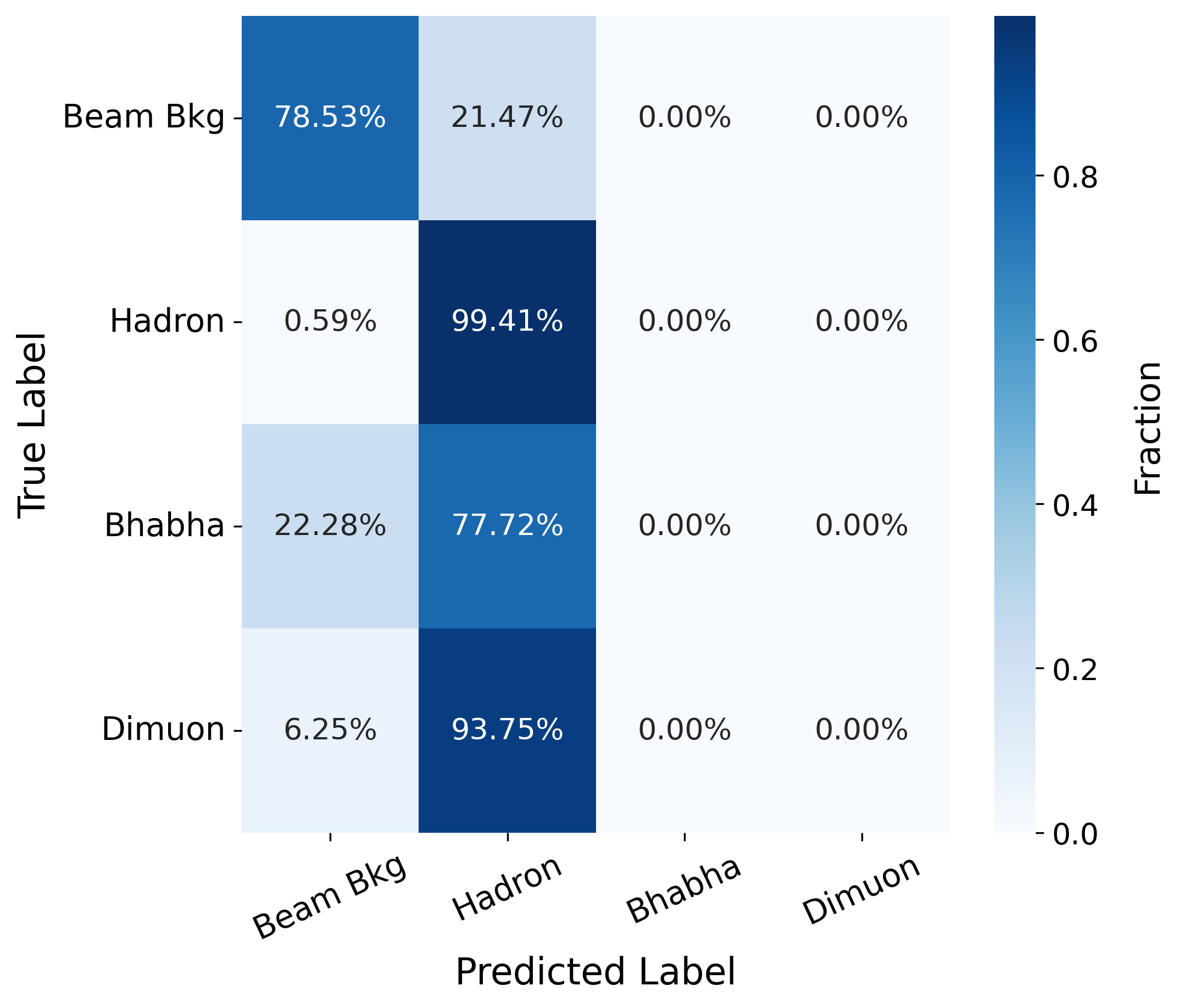}
\caption{EMC only}
\label{fig:matrix_four_nn_4680_emc}
\end{subfigure}
\caption{Normalized confusion matrices at $\sqrt{s}=4.680~\mathrm{GeV}$ for
the two feature sets.}
\label{fig:4680_four_matrix}
\end{figure*}

Figure~\ref{fig:4680_four_matrix} shows the resulting normalized confusion
matrices. Table~\ref{tab:4680_four_perf_diff} gives the corresponding
performance comparison.

\begin{table}[htbp]
\centering
\caption{Performance of the high-energy model relative to the
traditional method at matched hadron-event efficiency.}
\label{tab:4680_four_perf_diff}
\resizebox{\columnwidth}{!}{%
\begin{tabular}{ccccc}
\toprule
\textbf{Method} & \multicolumn{2}{c}{\textbf{MDC\&TOF only}} &
\multicolumn{2}{c}{\textbf{EMC only}} \\
\midrule
& \textbf{TPR (\%)} & \textbf{FPR (\%)} &
\textbf{TPR (\%)} & \textbf{FPR (\%)} \\
\midrule
Traditional method & 93.82 & $60.53\pm0.03$ & 99.41 & $53.14\pm0.02$ \\
NN Model & -- & $48.86\pm0.03$ & -- & $21.47\pm0.01$ \\
\midrule
$\mathbf{R}_{\mathbf{FPR}}$ &
\multicolumn{2}{c}{$\mathbf{19.28\%}$} &
\multicolumn{2}{c}{$\mathbf{59.60\%}$} \\
\bottomrule
\end{tabular}
}
\end{table}

\begin{figure*}[htbp]
\centering
\begin{subfigure}[b]{0.90\textwidth}
\centering
\includegraphics[width=\textwidth]{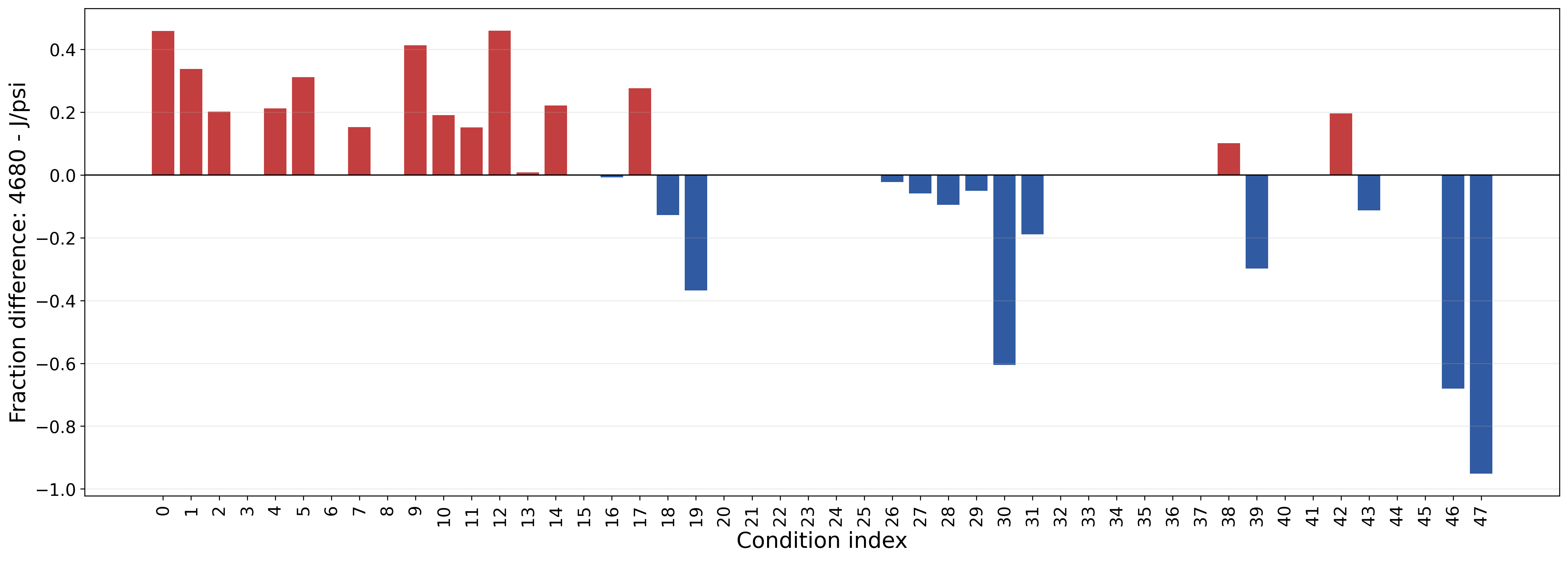}
\caption{Beam-background difference}
\end{subfigure}

\par\vspace{5pt}

\begin{subfigure}[b]{0.90\textwidth}
\centering
\includegraphics[width=\textwidth]{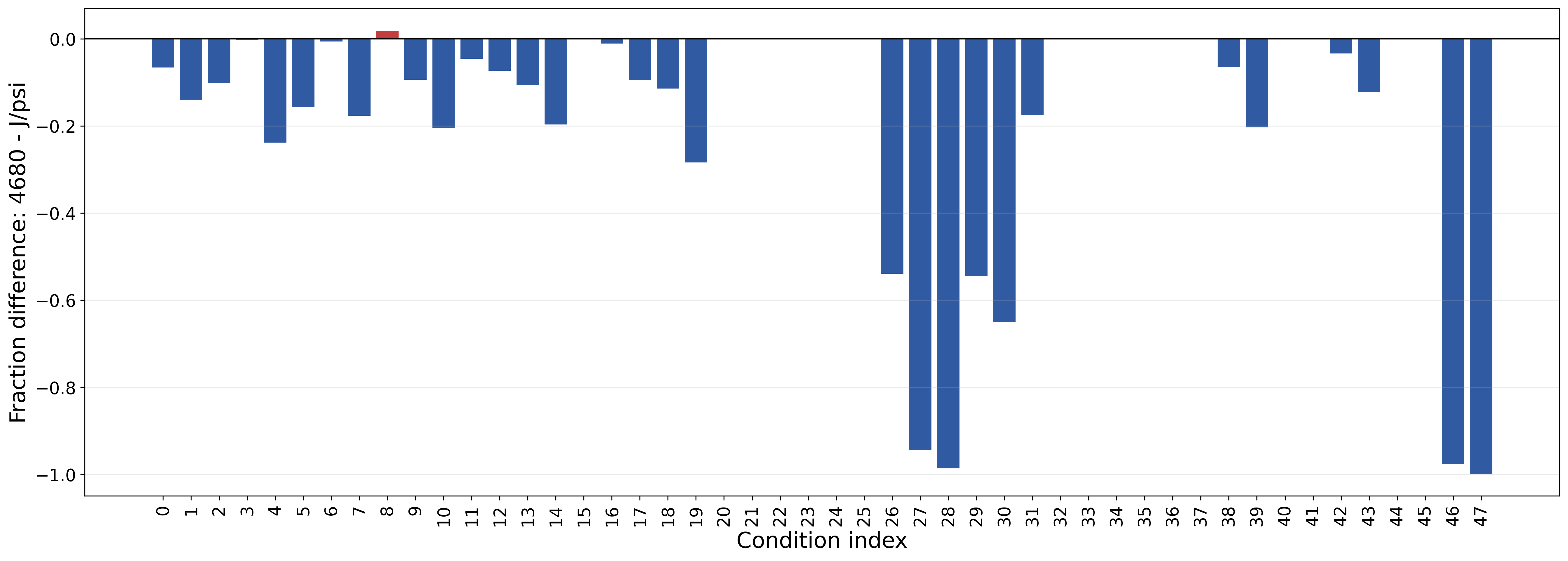}
\caption{Hadron-event difference}
\end{subfigure}

\caption{Differences between the mean input-condition values at
$\sqrt{s}=4.680~\mathrm{GeV}$ and $\sqrt{s}=3.097~\mathrm{GeV}$ for
(a) beam-background and (b) hadron events.}
\label{fig:feature_diff}
\end{figure*}

The results at $\sqrt{s}=4.680~\mathrm{GeV}$ differ significantly from those
at $\sqrt{s}=3.097~\mathrm{GeV}$. To investigate this difference, the input-feature distributions at the two energy points are compared. For each feature, its mean over all events is computed. The quantity
$\Delta\bar{\mathbf{c}}=\bar{\mathbf{c}}_{4.680}-
\bar{\mathbf{c}}_{3.097}$ is then evaluated, where $\bar{\mathbf{c}}$ denotes the vector of mean
input-condition values. The difference is evaluated separately for
beam-background and hadron events to identify the dominant sources of the
discrepancy. Figure~\ref{fig:feature_diff} shows larger differences for the
hadron-event features. These distributional differences likely contribute to
the observed variation in model performance.

\subsection{Random-Trigger Generalization}
\label{sec:random-trigger-test}
\begin{figure}[htbp]
    \centering
    \includegraphics[width=\columnwidth]{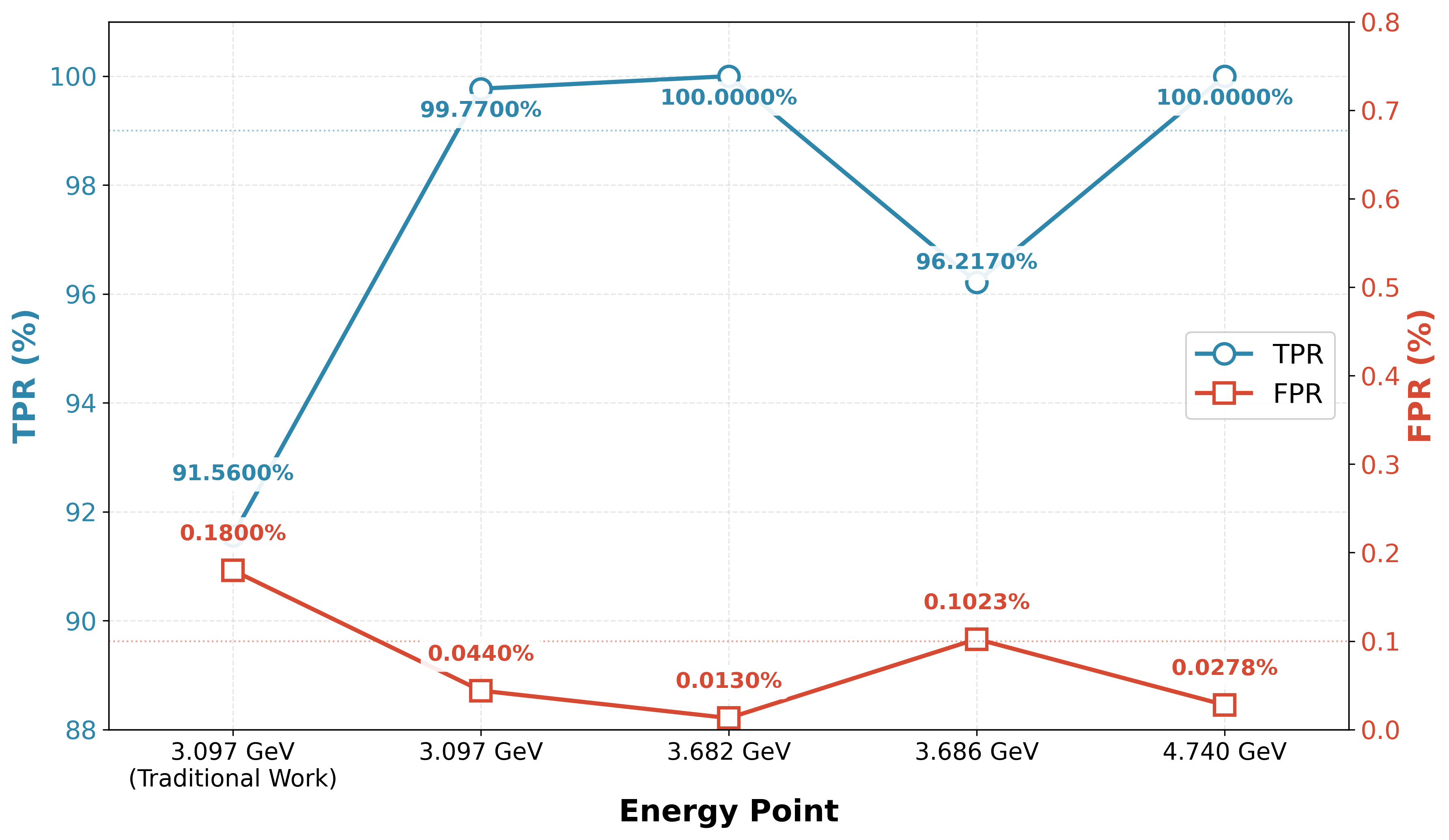}
    \caption{\raggedright Hadron-event TPR and beam-background FPR in random-trigger data across center-of-mass energies.}
    \label{fig:random_test}
\end{figure}

The random-trigger data were collected with only Channel~10 enabled and are therefore independent of the physics-trigger conditions. Most events in these samples are beam-background events. Their energy points and event counts are listed in the random-trigger block of Table~\ref{tab:data_samples}. These data are used to test whether the classifier retains its beam-background rejection capability for samples collected at energies not represented in the training data. At each energy point, random-trigger events are combined with hadron events from the \(\sqrt{s}=3.097~\mathrm{GeV}\) sample. The model trained at \(\sqrt{s}=3.097~\mathrm{GeV}\) is then applied without retraining. The energy points are \(\sqrt{s}=3.097\), \(3.682\), \(3.686\), and \(4.740~\mathrm{GeV}\). Figure~\ref{fig:random_test} shows the results and the performance of the established trigger selection on the \(\sqrt{s}=3.097~\mathrm{GeV}\) random-trigger sample as a reference. At all tested energies, the classifier maintains high hadron-event efficiency while suppressing a substantial fraction of beam background. At \(\sqrt{s}=3.097~\mathrm{GeV}\), its beam-background rejection is compared with that of the established selection using the same random-trigger sample.

\section{HARDWARE IMPLEMENTATION FEASIBILITY}
\label{sec:hardware}
The hardware performance of the model is first evaluated through hardware-oriented synthesis with \texttt{hls4ml}~\cite{fastml_hls4ml,Duarte:2018ite,Schulte:2025mai}. The target device is an AMD Xilinx Virtex UltraScale+ XCVU9P-FLGA2104-3-E FPGA. The current synthesis result gives an estimated inference latency of \(35~\mathrm{ns}\) at a clock frequency of \(200~\mathrm{MHz}\), and the overall resource utilization is below 50\% for all reported FPGA resource categories. Detailed simulation results are presented in Table~\ref{tab:fpga_performance}.

These results provide an initial indication of the model's hardware feasibility. The present synthesis does not include the timing associated with the trigger-selection logic or its integration with the complete trigger chain. Consequently, the estimated latency and resource utilization can differ from those obtained in a deployed system. Further hardware studies remain necessary, including simulations that incorporate realistic trigger timing and implementation conditions. Reliable estimates of hardware latency and resource utilization require validation in representative operating scenarios. These hardware tests will be progressively refined in subsequent work.

\begin{table*}[htbp]
\centering
\caption{Latency and FPGAs resource utilization of the implemented model.}
\label{tab:fpga_performance}
\begin{tabular*}{\textwidth}{@{\extracolsep{\fill}}ccc cccc@{}}
\toprule
 \textbf{Implementation} & \multicolumn{2}{c}{\textbf{Latency}}
 & \multicolumn{4}{c}{\textbf{Resource utilization}} \\
 & \textbf{Cycles} & \textbf{Absolute}
 & \textbf{DSP48E} & \textbf{LUT} & \textbf{FF} & \textbf{BRAM$_{18\mathrm{K}}$} \\
\midrule
Optimized model
& 7
& $35\,\mathrm{ns}$
& 13 ($0.8\%$)
& 99,838 ($10.8\%$)
& 21,784 ($1.2\%$)
& 3 ($<0.1\%$) \\
\bottomrule
\end{tabular*}
\end{table*}

\section{CONCLUSION}
\label{sec:conclusion}
High-luminosity lepton colliders require trigger strategies that control beam-induced backgrounds while preserving their broad physics programmes. 
Experimentally recorded BESIII data, rather than simulated backgrounds, are used to train and evaluate compact neural-network classifiers for trigger-level beam-background suppression. 
The classifiers augment the established binary trigger logic and provide additional beam-background rejection while preserving signal-trigger efficiency. 
At matched hadron-event TPR, the four-class model reduces the beam-background FPR by 20.38\% for the MDC\&TOF-only sample and by 31.16\% for the EMC-only sample at $\sqrt{s}=3.097~\mathrm{GeV}$. 
At higher center-of-mass energies, where beam backgrounds are more prominent, the corresponding reductions are 19.28\% and 59.60\% for the MDC\&TOF-only and EMC-only samples, respectively. 
Tests on random-trigger data confirm that the model maintains its beam-background rejection capability across the studied energy range. 
These results indicate that compact neural-network classifiers can serve as a complementary option in future BESIII trigger upgrades and provide a reference for trigger optimization at future high-luminosity lepton colliders.

The use of experimental data captures the dependence of beam background on the actual machine state and complements simulations of time-dependent accelerator conditions. Early identification of beam background can support the efficient use of front-end buffers, DAQ resources, and downstream reconstruction. The approach is relevant to future operating conditions with increased beam-induced background rates. The present results use the existing binary trigger-condition inputs; future front-end upgrades could provide lower-level and more granular detector information for subsequent studies.

Hardware-oriented synthesis provides an initial indication that FPGA deployment of the compact multilayer NN is feasible. The estimated inference latency is \(35~\mathrm{ns}\), with overall resource utilization below 50\%. Because the present study does not yet include the complete trigger-selection timing, these values require validation in representative operating scenarios. The study provides a data-driven framework for future trigger upgrades and a relevant reference for high-luminosity \(e^+e^-\) facilities, including the STCF, CEPC, and FCC-ee~\cite{stcf_l1_trigger,cepc_cdr,fcc_ee_physics}.

\section{ACKNOWLEDGEMENTS}
\label{sec:acknowledgements}
The authors gratefully acknowledge the support of BESIII and BEPCII. 
This work is supported by the Strategic Priority Research Program of the Chinese Academy of Sciences under Grant No. XDA0480600,
the National Natural Science Foundation of China under Grant Nos. 12405099 and 12675104,
and the Fundamental Research Funds for the Central Universities under Contract No. 010-63263110.

\end{document}